\documentclass[12pt]{article}
\usepackage{amssymb,amsmath,amsfonts,epsfig,graphics}

\def\l{\left(}
\def\r{\right)}

\newcommand{\be}{\begin{equation}}
\newcommand{\ee}{\end{equation}}
\newcommand{\bea}{\begin{eqnarray}}
\newcommand{\eea}{\end{eqnarray}}

\newcommand{\bg}{\begin{gather}}
\newcommand{\eg}{\end{gather}}

\begin{document}

\title{Invisible Higgs in weak bosons associative production with
  heavy quarks at LHC: probing mass and width
}
\author{
E.~E.~Boos\thanks{{\bf e-mail}: boos@theory.sinp.msu.ru}, 
\\
{\small{\em
Skobeltsyn Institute of Nuclear Physics, Moscow State University, }}\\
{\small{\em
Vorobiovy gory, Moscow 119991, Russia
}}\vspace{0.3cm}\\ 
S.~V.~Demidov\thanks{{\bf e-mail}: demidov@ms2.inr.ac.ru} 
and 
 D.~S.~Gorbunov\thanks{{\bf e-mail}: gorby@ms2.inr.ac.ru}  
\\
{\small{\em
Institute for Nuclear Research of the Russian Academy of Sciences, }}\\
{\small{\em
60th October Anniversary prospect 7a, Moscow 117312, Russia
}}
}
\date{}
\maketitle
 \begin{abstract}
New physics coupled to the Higgs boson may hide it in the standard
decay channels to be investigated at LHC. We consider the models where
new invisible dominant decay modes of the Higgs boson are responsible
for this hiding. We propose to study at LHC the weak boson production
associated with heavy quarks: our analysis revealed that boson pair
invariant mass distribution is sensitive to both mass and width of the
invisible Higgs boson, if it is not too far from the weak boson pair 
threshold. We present tree-level results for the most relevant cases of
top quarks and of bottom quarks in Standard Model extensions with
large $b$-quark Yukawa coupling. We argue that QCD
corrections do not spoil these results allowing for unambiguous
extraction of the Higgs boson mass and width from the analysis of 
large enough amount of data. 
 \end{abstract}

\section{Introduction}
\label{Sec:intro}

The main task to accomplish at LHC is searching for the Higgs boson,
the only particle of the Standard Model (SM) has not been observed so 
far. However, it has become clear that the SM itself is
incomplete. There are neutrino oscillations, strong CP-problem, dark
matter, baryon asymmetry of the Universe which lack for explanation 
within the SM (see, e.g. review section of
Ref.~\cite{Amsler:2008zzb}).  

In the SM only the Higgs sector remains still hidden, while all others 
were thoroughly explored without any convincing direct evidence of new
physics processes in there besides neutrino oscillations. So, it can
happen that at LHC a new physics will show up (for the first time?)
right in the Higgs sector. Generally, in SM extensions both the Higgs
boson production and decay rates deviate significantly from the SM 
prediction. May be the simplest yet physically-motivated extension of
this type is the model of dark matter scalar particles 
\cite{McDonald:1993ex+Burgess:2000yq}. This extension contains one
new scalar field singlet under the SM gauge group. Parity ensures that 
the new scalar particle is stable and hence the dark matter
candidate. The only renormalizable interaction of this scalar 
with SM particles is coupling to the SM Higgs field~\footnote{In this
  type of models the new scalar can also couple to other new
  (otherwise hidden) fields, e.g. can transform non-trivially under
  some new (hidden) gauge group. In literature models of this kind are
  dubbed as {\it Higgs portal }~\cite{Patt:2006fw}, as the only door
  to the hidden fields is in the Higgs sector. }.  This coupling is
responsible for the dark matter generation in the early Universe due
to Higgs boson scatterings in primordial plasma. Thus, the 
(virtual) SM Higgs boson can decay into two dark matter 
particles. Appearance of this {\it invisible } (real or virtual) decay
mode is the only tree-level modification to be searched for at LHC in
this simple model.

There are many other physically-motivated and (much) more complicated
extensions of the SM where properties of the Higgs boson get
modified~\cite{Shrock:1982kd}. Among them are models with additional
spatial dimensions suggesting a solution to the problem of hierarchy
between the electroweak and Plank scales
\cite{ArkaniHamed:1998rs,Randall:1999ee}. There are also multi-Higgs
models capable of explaining the baryon asymmetry of the Universe, see
e.g. \cite{Fromme:2006cm}. Supersymmetric extensions of the SM address
both problems and also have (natural) candidate(s) to be dark matter
particles. In all the models above, as well as in many other
extensions of the SM, the Higgs boson starts to interact with new
fields. This changes its production rates and decay pattern to an
extent which in some cases (see,
e.g.~\cite{Shrock:1982kd,Belotsky:2002ym,Krasnikov:1997nh,Hosotani:2008tx}) 
becomes crucial for the prospects of a particular experiment in
searches for the Higgs boson.   

Let us consider a class of such extensions of the SM where new
invisible Higgs decay mode dominates. In these models the branching
ratios of all visible decay modes of the Higgs boson get decreased,
sometimes to the level beyond the LHC sensitivity (see, e.g., discussion
in Refs.~\cite{Binoth:1996au,Bock:2010nz}). The question is how to
search for the Higgs boson and measure its major parameters mass
and width in this case, where the Higgs boson is invisible itself.

The strategy for hunting the invisible Higgs boson at LHC has been
developed in literature to a certain extent. In particular, the Higgs
boson production with subsequent decay to invisible mode gives rise to
missing $P_T$ signature in various channels such as Vector Boson
Fusion $qq\to qqH$~\cite{Eboli:2000ze} and the associated production
processes, $gg\to t\bar{t}H$~\cite{Gunion:1993jf} and $qq\to ZH$ or
$qq\to W^{\pm}H$~\cite{Choudhury:1993hv,Godbole:2003it}.  A tricky
question here is how to make sure that the observed signal is really
due to production of the Higgs boson, not some other particle.
Another disadvantage of the missing $P_T$ signature is that only the
Higgs boson mass can be estimated from the data analysis: the Higgs
boson width remains unobservable.

To address both issues one can consider the very channels where the
exchange of the Higgs boson restores unitarity in SM particle
collisions. These are channels with massive vector bosons in initial
and/or final states 
\cite{Lee:1977eg,Chanowitz:1978mv,Appelquist:1987cf}.  The relevant
processes at LHC are inclusive $ZZ$ and $W^+W^-$ production. In case
of the Higgs boson mass above the weak boson pair thresholds these
processes have been thoroughly studied \cite{Bagger:1995mk}.  The
Higgs boson mass can be reconstructed, e.g., from analysis of invariant
mass of outgoing final leptons, while measurement of the inclusive
production rate yields the decay branching ratio to weak bosons. The
latter implies the estimate of the total Higgs boson width (saturated
by invisible mode) assuming the SM coupling between the Higgs and weak
gauge bosons.

In this paper we are interested mostly in the opposite case where the
Higgs boson decays to weak bosons are kinematically forbidden on
shell. This case of the light Higgs boson is certainly favorable from
the combined analysis of electroweak precision
measurements~\cite{Amsler:2008zzb,Alcaraz:2009jr}. We concentrate on
the models, where the Higgs boson width is dominated by the invisible
mode, hence the Higgs branching ratios into SM particles are
suppressed with respect to the estimates within the SM.  In this setup 
virtual Higgs boson contributions to $W^+W^-$ and $ZZ$ production via
weak boson fusion and gluon fusion have been already considered in
literature, see, e.g.,~\cite{Davoudiasl:2004aj}.

The main observation of this work is that the measurements of total
cross section and invariant mass distribution of the weak boson pair
generally allow to estimate {\it both the Higgs boson mass and
  width. }  This statement is illustrated by the realistic example of
the Higgs boson contribution to the weak boson production associated
with $t$-quarks: $pp\to ZZ\bar t t$, $pp\to WW \bar t t$.  Likewise we
study processes $pp\to ZZ\bar b b$ and $pp\to WW \bar b b$. The
production rates are too small here, but might be of some interest in
the SM extensions where the Higgs boson coupling to $b$-quark is
amplified, e.g., in multi-Higgs or supersymmetric models.  The 
dependence of the total cross section and the gauge boson invariant
mass on the Higgs boson parameters is explained analytically in this
example of associated production with $b\bar b$-pair. 

As the first step towards this goal we discuss in this paper only
signal properties, leaving the so called {\it reducible background }
without any discussion. Thus, we consider tree level contribution of
Higgs boson and all other SM particles to the processes presented
above: production of two weak bosons and two heavy quarks in
proton-proton scattering. Meanwhile, we estimated the size of
possible uncertainties in the obtained results due to QCD quantum 
corrections and found them small. Finally, we expect that the
invariant mass of weak boson 
pairs in other relevant channels, like $ZZ$ and $W^+W^-$ or $jjZZ$ and
$jj W^+W^-$ is also sensitive to both the SM Higgs boson mass and
width, but leave the corresponding study for future work. Note that
the dependence of the invariant mass on the Higgs mass in these
channels has been considered in literature (see,
  e.g.,~\cite{Davoudiasl:2004aj}).

As a side remark, come back to the fundamental role of the Higgs boson
in particle scatterings, which exchange restores unitarity in the
electroweak theory. To the weak boson production associated with heavy
quarks we are considering here, both gauge and Yukawa Higgs couplings 
contribute. Hence, its study allows to test the Higgs mechanism both
in the gauge and in the fermion sectors. This is an important task and
it is worth to be performed at LHC and SLHC even if the Higgs boson
signal would closely resemble what SM predicts.
 
The paper is organized as follows. In Section~\ref{Sec:tt} we discuss
the weak boson pair production associated with top quarks --- the
process which study at LHC gives a chance to measure both the width
and mass of the invisible Higgs boson. We consider in
Section~\ref{Sec:bb} similar processes with $b$-quarks in models where
$b$-quark Yukawa coupling is amplified with respect to that of in the
SM. Simple analytical formulas presented in Section~\ref{Sec:analytic}
for weak boson production in partonic scatterings of quarks illustrate
the obtained results and provide with understanding of the physics
behind. Section~\ref{Sec:conclusion} contains summary.  

\section{Invisible Higgs in $pp\to t\bar{t}ZZ$ 
and $pp\to t\bar{t}W^+W^-$ at LHC}
\label{Sec:tt}

We begin our study with the channel  $pp\to t\bar{t}ZZ$. We would
like to consider virtual contribution of the Higgs boson to these
processes, since it remains almost the same as in the SM. At the same
time, as it is mentioned in Introduction, a large rate of invisible
Higgs decay would prevent searches for the Higgs boson by means of 
usual methods, that is by studying Higgs boson decay modes. In these
final states the Higgs contribution will be considerably suppressed
by the ratio of the Higgs boson width to that within the SM. 

We use CompHEP~\cite{Pukhov:1999gg,Boos:2004kh,cite:comphep} to
calculate the tree level partonic cross sections of the process 
$pp\to t\bar{t}ZZ$. This process is dominated mostly by the
subprocess $gg\to t\bar{t}ZZ$, where the virtual Higgs boson decays
into $ZZ$. We note, that both signal (Higgs boson contribution) and SM
irreducible background amplitudes are taken into account in our
analysis. We adopt CTEQ6L1~\cite{Pumplin:2002vw} for parton
distribution function in our calculations. The width of Higgs boson is
calculated by HDECAY program~\cite{Djouadi:1997yw}. No phase space
cuts are imposed on the final state. In this paper we outline general
behavior of the cross section and $ZZ$ invariant mass distribution and
postpone detailed calculations, including analysis of the reducible
background and the detector response, for future work. It is worth to
note in this respect that dealing with a final state of several heavy
particles one can hope to sufficiently suppress the reducible
backgrounds playing with cuts on the relevant physical observables.

\begin{figure}[!htb]
\begin{center}
\begin{tabular}{ll}
\includegraphics[angle=0,width=0.50\columnwidth]{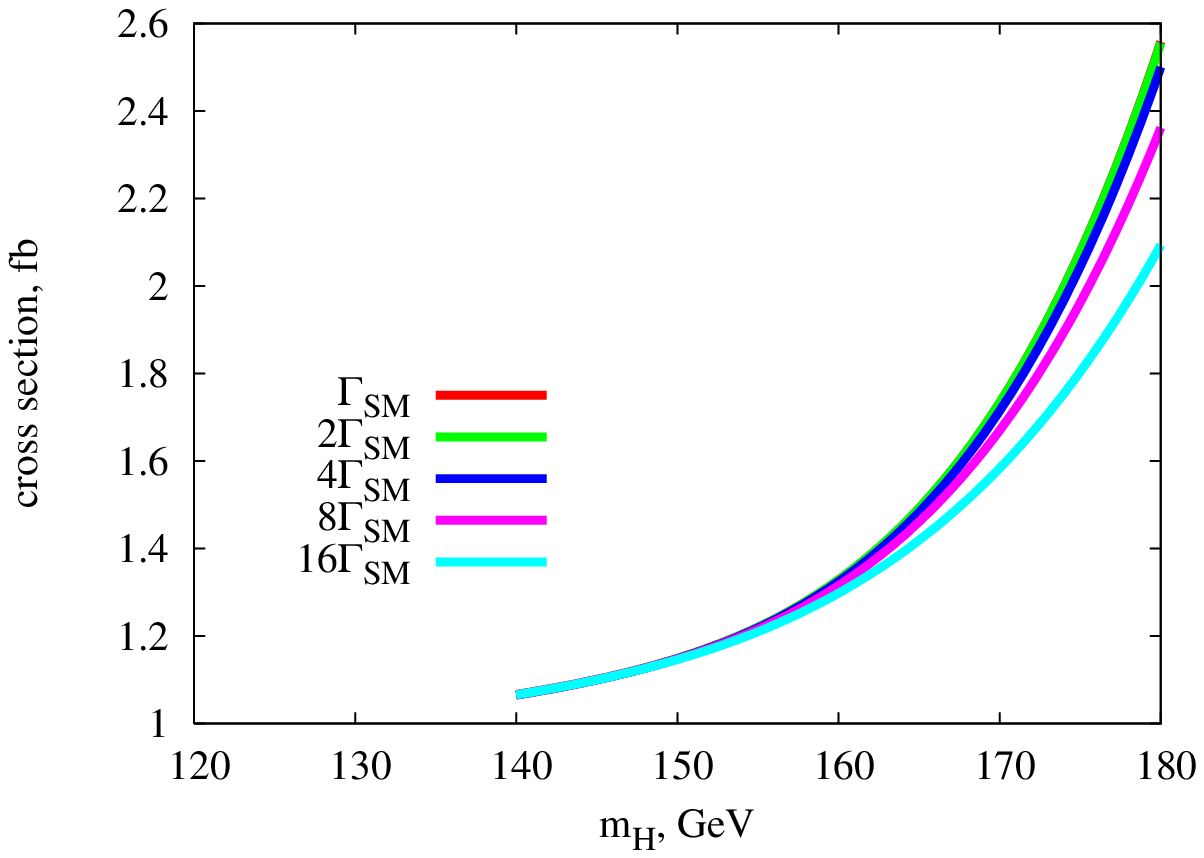} 
&
\includegraphics[angle=0,width=0.50\columnwidth]{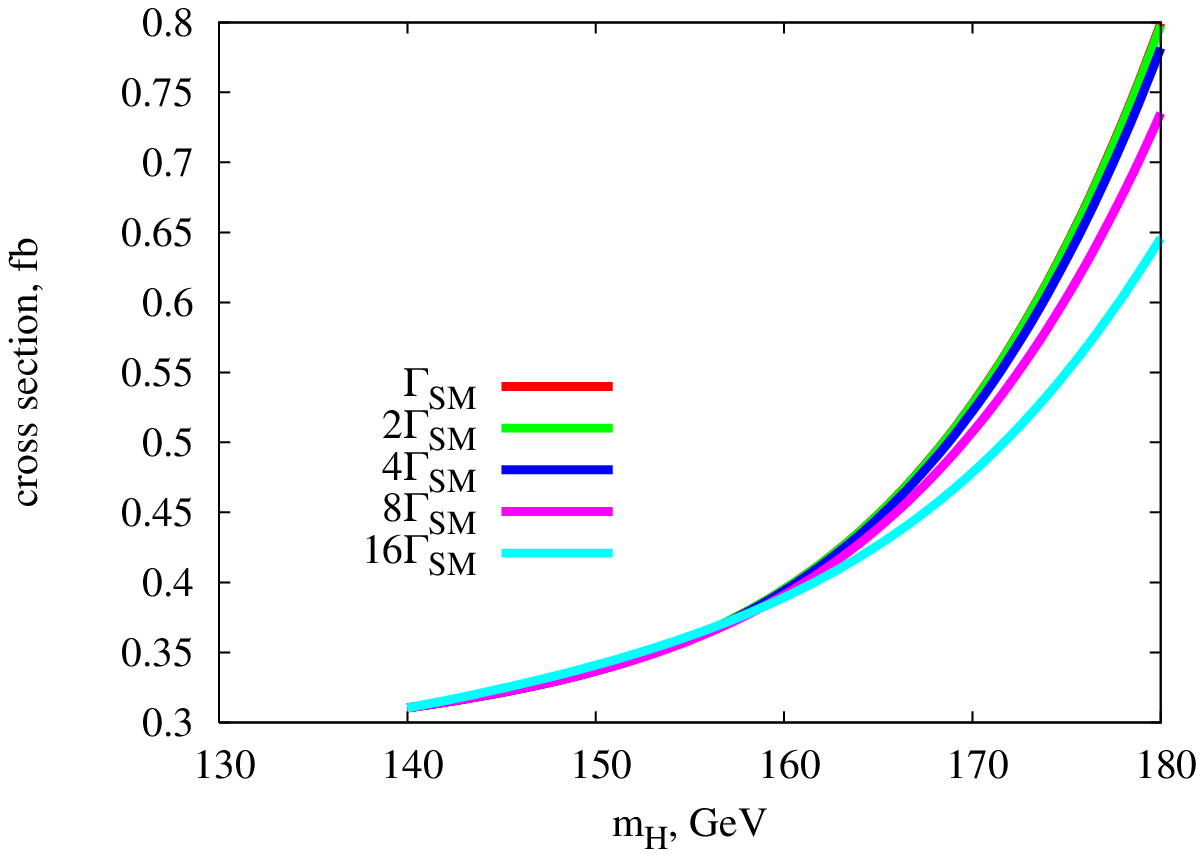} 
\end{tabular}
\end{center}
\caption{\label{cs} The dependence of the total cross section
  $pp\to t\bar{t} ZZ$ at $\sqrt{s}=14$~TeV (left panel) and
  $\sqrt{s}=10$~TeV (right panel) on the mass of the Higgs boson in
  for a set of values of the Higgs boson width. Here $\Gamma_{SM}$ is
  the width of the Standard Model Higgs boson. 
}
\end{figure}

\begin{figure}[!htb]
\includegraphics[angle=0,width=0.80\columnwidth]{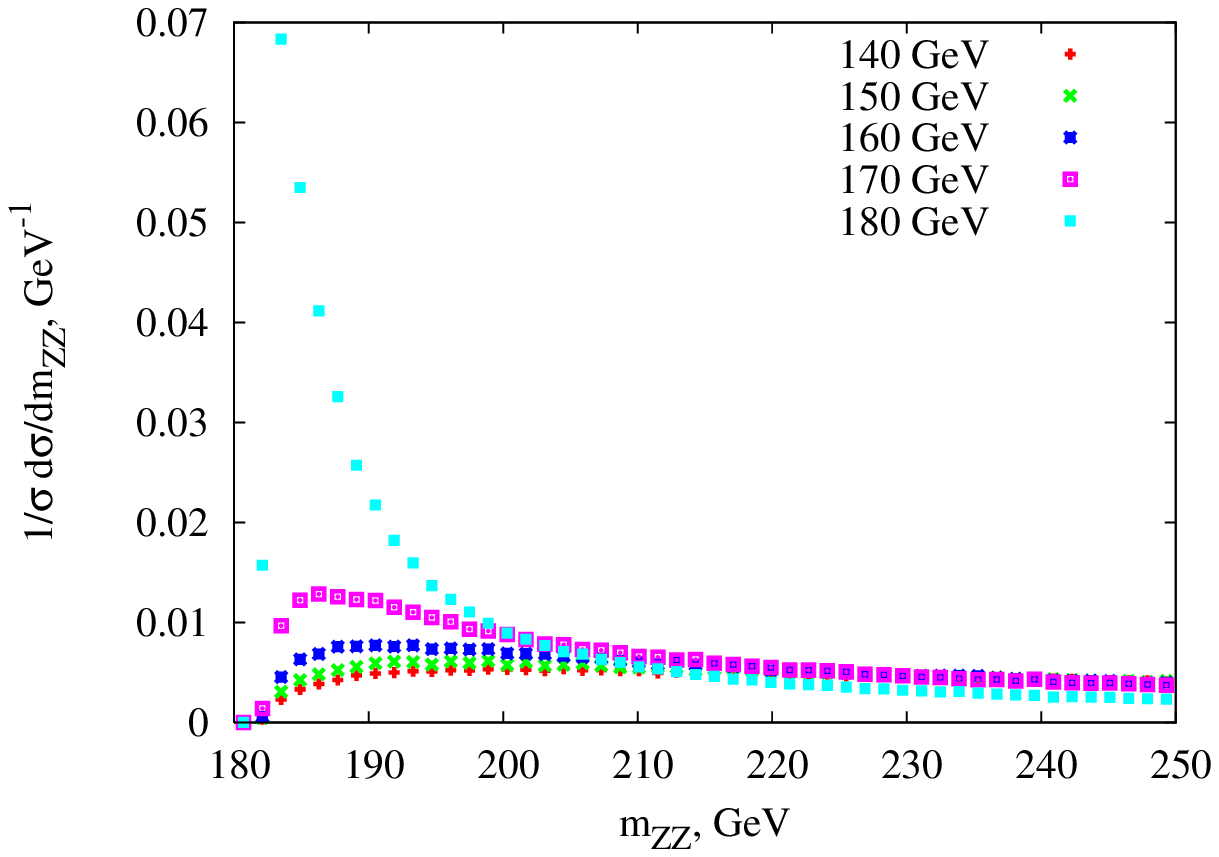} 
\includegraphics[angle=0,width=0.80\columnwidth]{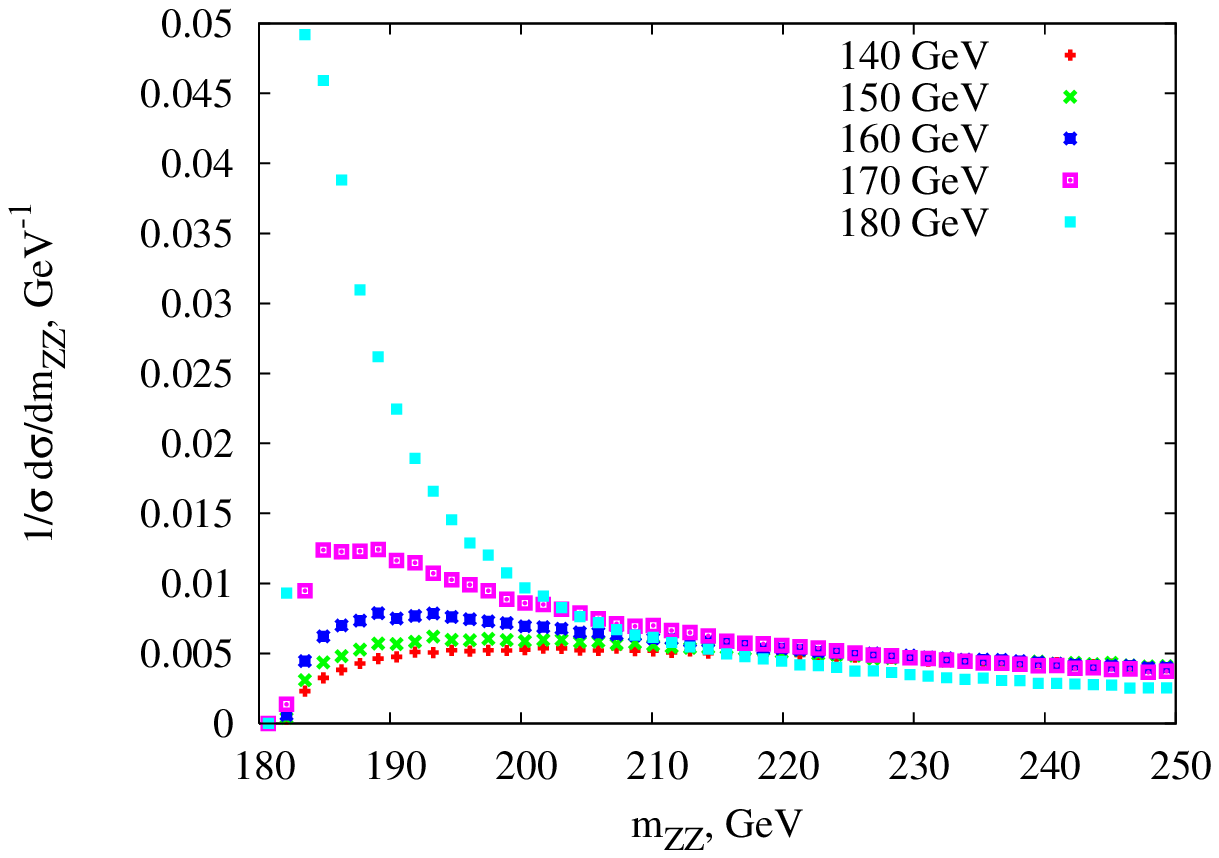} 
\caption{\label{ZZ} The invariant mass $m_{ZZ}$ distribution for
  the $pp\to t\bar{t} ZZ$ at $\sqrt{s}=14$~TeV process for several
  values of the Higgs boson mass for Standard Model Higgs width (upper 
  panel) and for the width which is 8 times larger (lower panel). 
}
\end{figure}

\begin{figure}[!htb]
\includegraphics[angle=0,width=0.8\columnwidth]{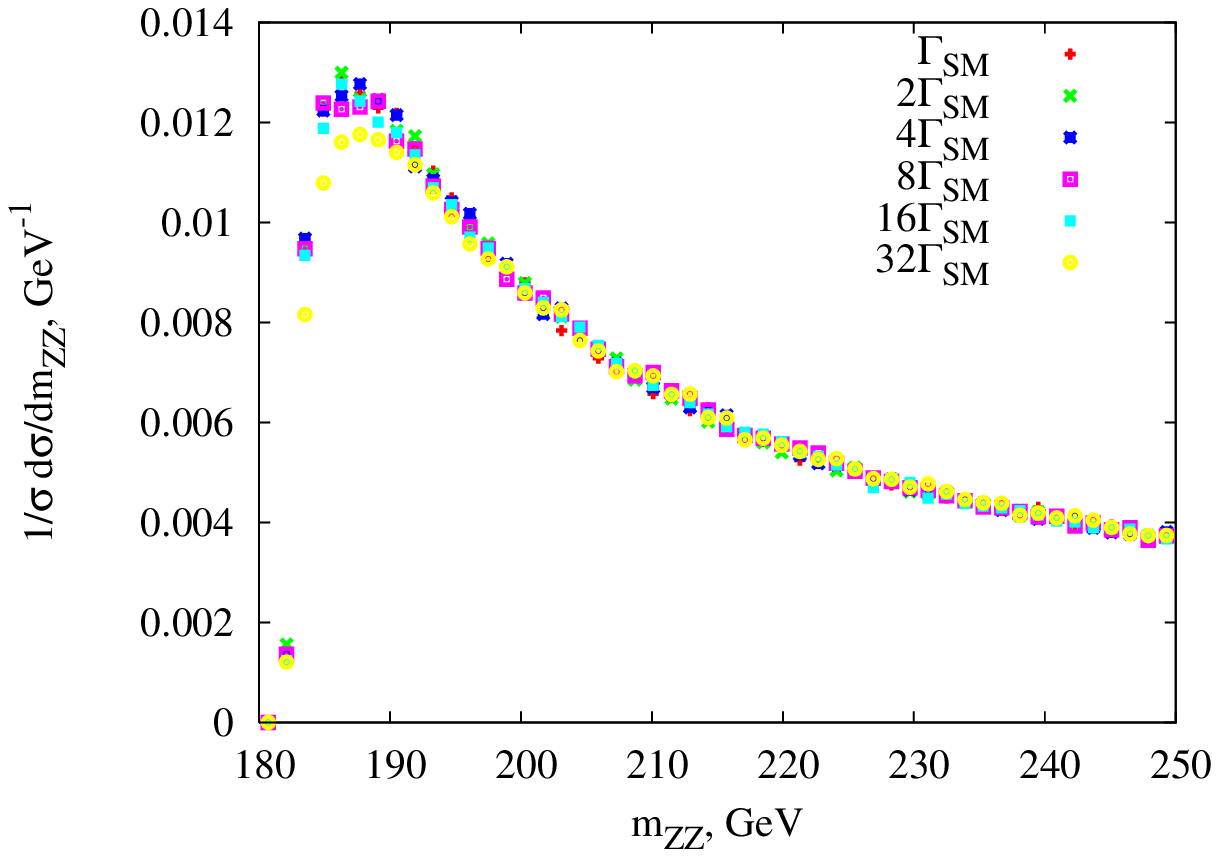} 
\includegraphics[angle=0,width=0.8\columnwidth]{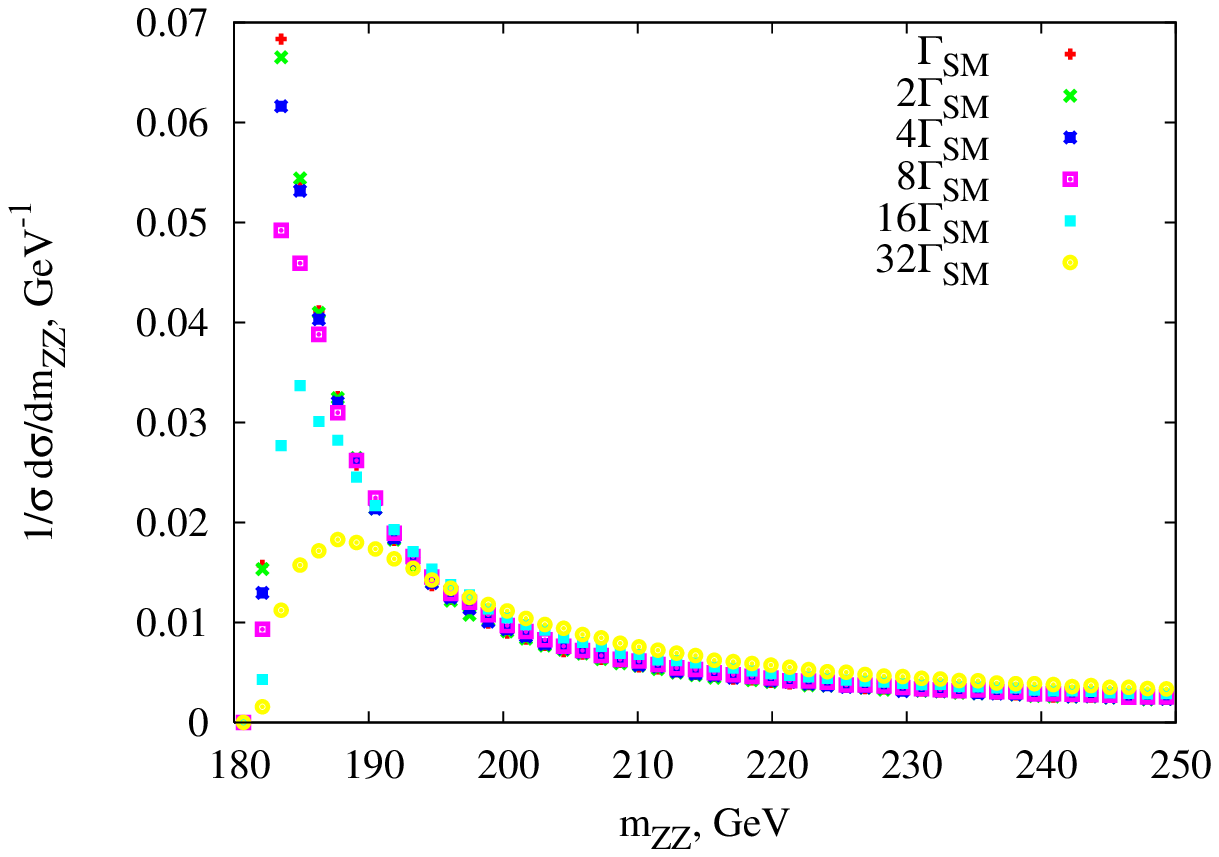} 
\caption{\label{width_ZZ} The dependence of the $m_{ZZ}$ invariant mass
  distribution in  $pp\to t\bar{t} ZZ$ on the width of the Higgs boson
  for the following values of masses $m_{H}$: $170$~GeV (top panel),
  $180$~GeV (bottom panel), at $\sqrt{s}=14$~TeV.
}
\end{figure}

\begin{figure}[!htb]
\begin{center}
\includegraphics[angle=0,width=0.80\columnwidth]{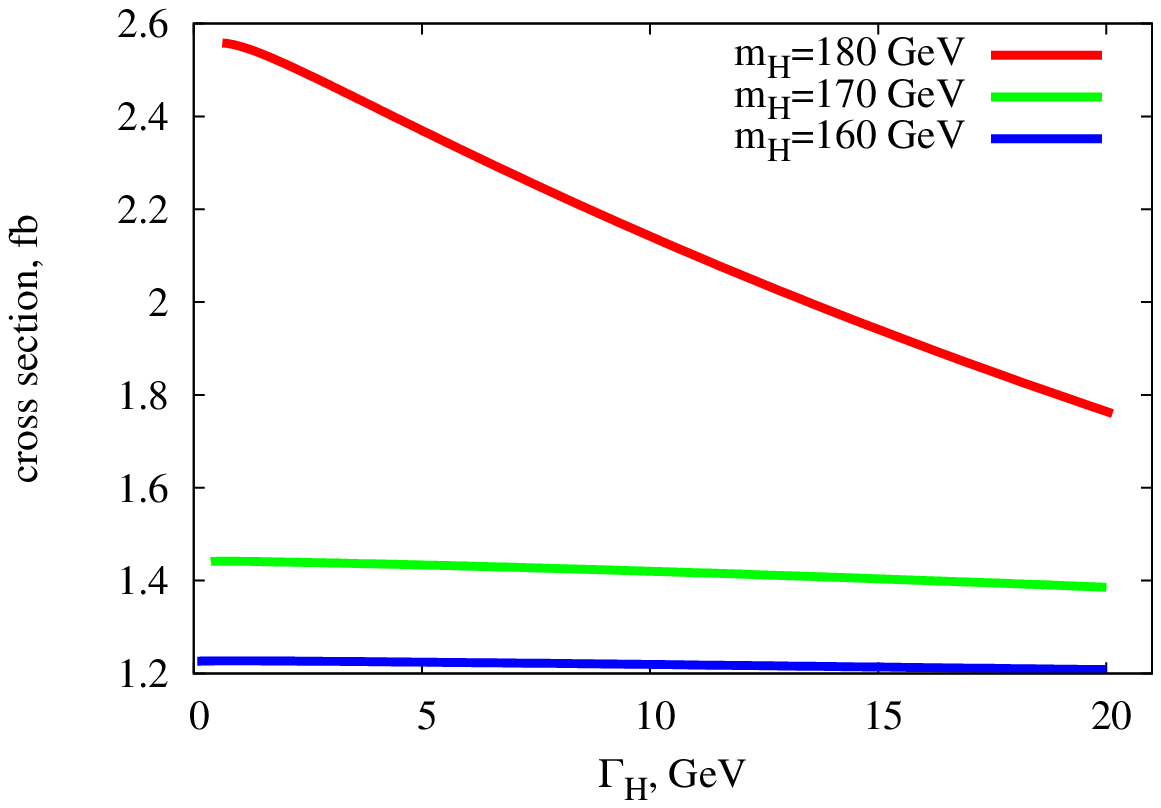} 
\end{center}
\caption{\label{width} The dependence of the total cross section
  $pp\to t\bar{t} ZZ$ at $\sqrt{s}=14$~TeV on the width of the Higgs
  boson for three values of masses $m_{H}$. 
}
\end{figure}

The results for total cross sections and invariant mass $Z$-boson pair
distributions are presented in Figures~\ref{cs}-\ref{width}. Here and
further we present results for collision energies of $\sqrt{s}=14$~TeV and
$\sqrt{s}=10$~TeV, for comparison. In Figure~\ref{cs} 
we plot the dependence of the total cross section of $pp\to
t\bar{t}ZZ$ channel at $\sqrt{s}=14$~TeV and $\sqrt{s}=10$~TeV on the
Higgs boson mass $m_H$ for a set of values of the Higgs boson width
$\Gamma_H$.  Here and further we perform study of the low mass range
of the Higgs boson mass preferred by the combined fit to electroweak 
precision measurements~\cite{Amsler:2008zzb,Alcaraz:2009jr}. As one 
expects,  at large values of the Higgs boson width the virtual Higgs
boson contribution to the amplitude of this process decreases and
hence the total cross section also decreases. Figure~\ref{ZZ} 
shows the corresponding invariant mass $m_{ZZ}$ distribution for
different values of Higgs boson mass in the cases of the SM Higgs
boson width (upper panel) and eight times larger width (lower panel).  
We see that both shape and position of maximum of
$m_{ZZ}$ distribution strongly depend on mass $m_{H}$, which can be 
used to pin down the Higgs boson mass. Moreover, from
Figure~\ref{width_ZZ}  
one concludes that this $m_{ZZ}$ distribution does not depend on the
width of the Higgs boson, except for the case of near threshold values
of its mass (see lower panel). Even in the latter case, for $m_{H}$
near $180$~GeV, we observe that the position of maximum in $m_{ZZ}$
distribution varies quite moderately with reasonable increase of the
Higgs boson width. 

Parameters of the Higgs boson --- mass and width --- can be obtained, 
as usual, from the combined two-parametric fit to the observables of
this channels. From the analysis above, however, we suggest the
following simple strategy for studying the invisible Higgs boson. One
measures $m_{ZZ}$ distribution and finds the mass of the Higgs boson
from the position of maximum in Figure~\ref{ZZ}. Then by using either
the dependence of the total cross section on the Higgs boson width
(see Figure~\ref{width})  
or (for the case of near threshold masses of the Higgs boson) the
shape of $m_{ZZ}$ distribution we determine $\Gamma_{H}$. 

\begin{figure}[htb]
\begin{center}
\begin{tabular}{ll}
\includegraphics[angle=0,width=0.50\columnwidth]{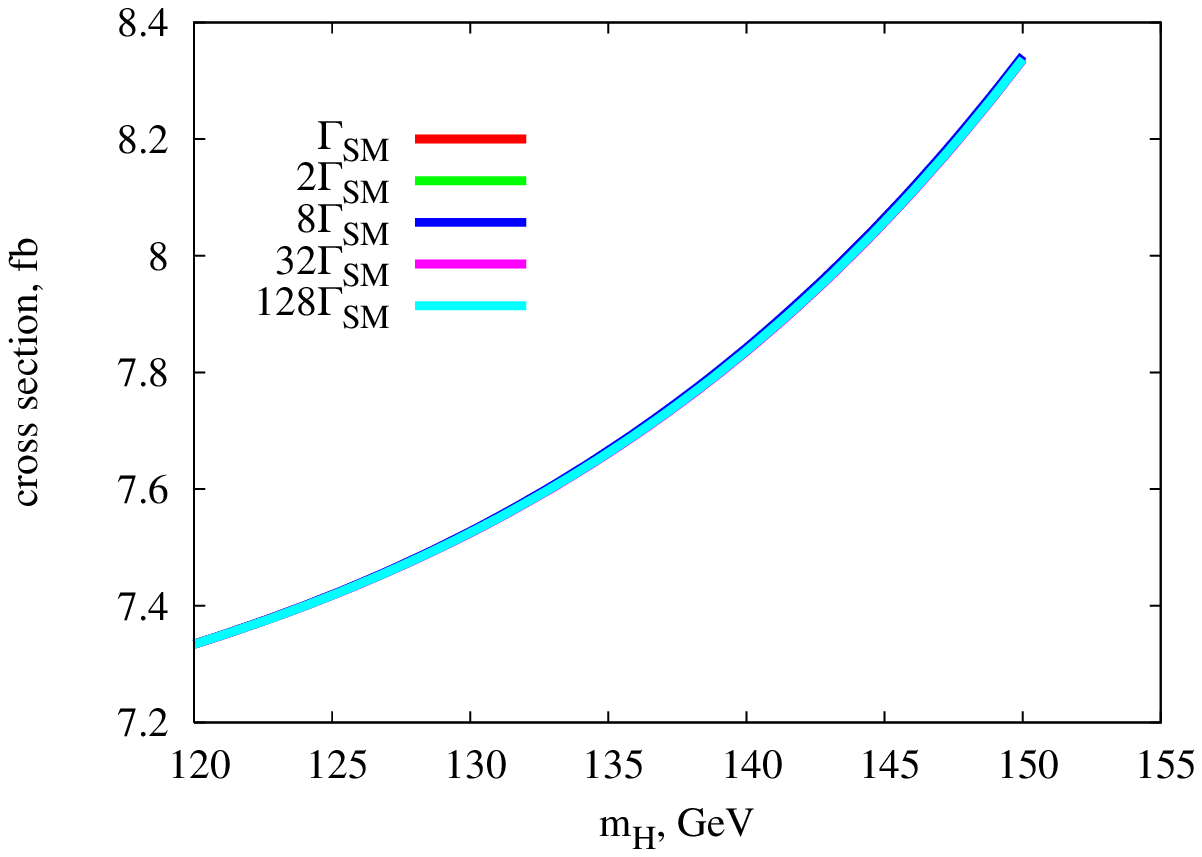} 
&
\includegraphics[angle=0,width=0.50\columnwidth]{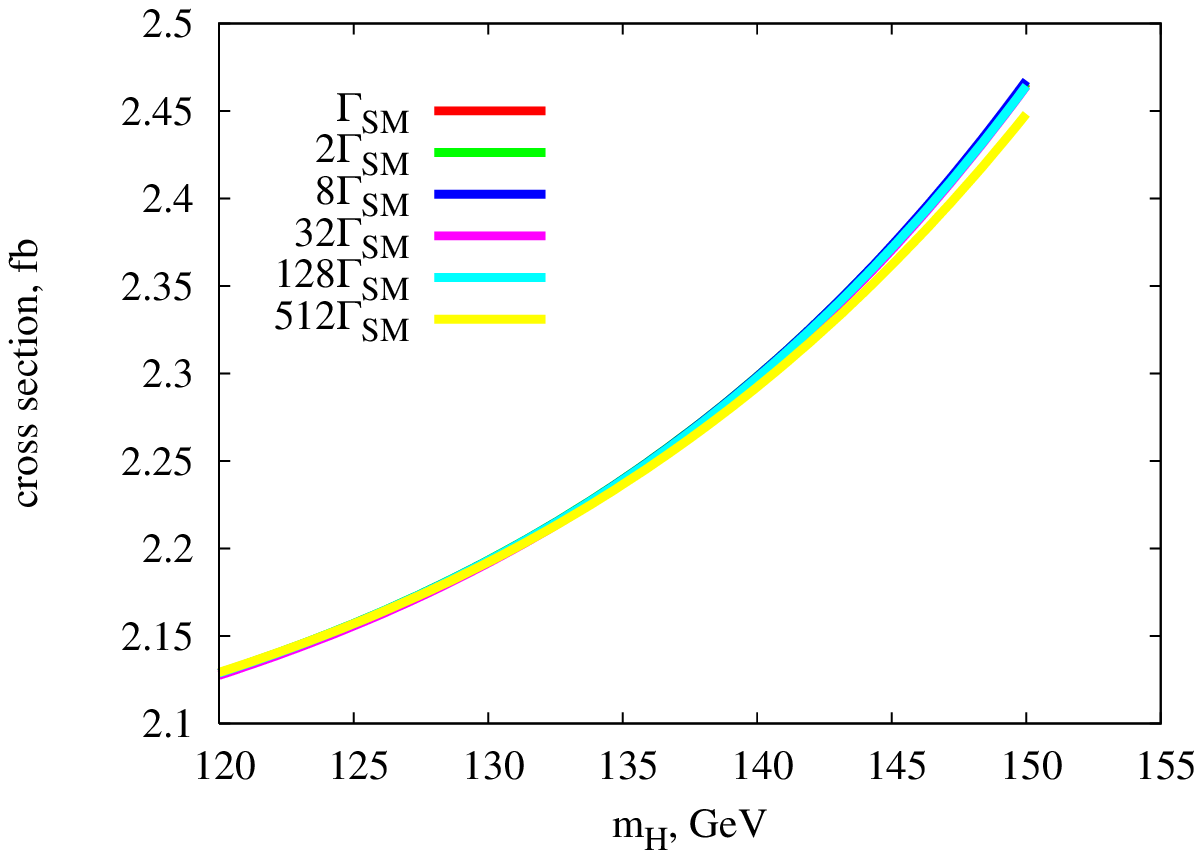} 
\end{tabular}
\end{center}
\caption{\label{cs_WW} The dependence of the total cross section
  of $pp\to t\bar{t} W^{+}W^{-}$ at $\sqrt{s}=14$~TeV (left) and
  $\sqrt{s}=10$~TeV (right) on the mass of the Higgs boson for various
  values of the Higgs boson width.
}
\end{figure}

\begin{figure}[htb]
\begin{center}
\includegraphics[angle=0,width=0.80\columnwidth]{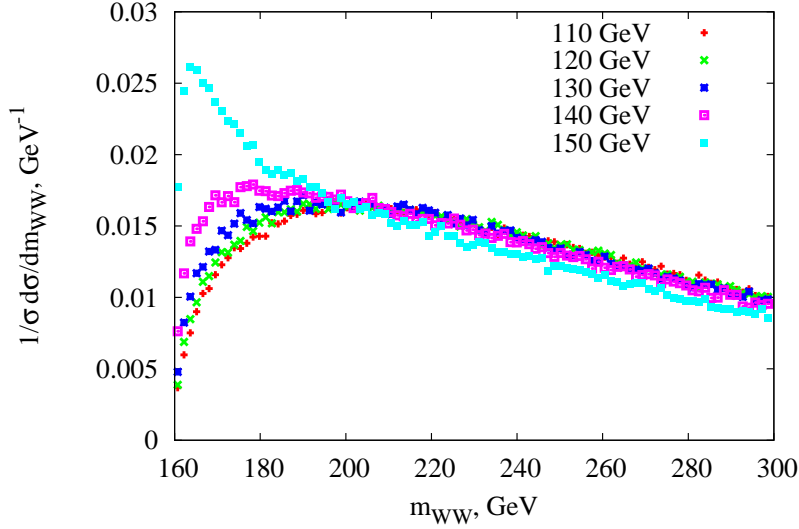} 
\end{center}
\caption{\label{WW} The invariant mass $m_{WW}$ distribution in
  $pp\to t\bar{t} W^{+}W^{-}$ at $\sqrt{s}=14$~TeV for
  several values of the Higgs boson mass. 
}
\end{figure}

\begin{figure}[htb]
\begin{center}
\includegraphics[angle=0,width=0.80\columnwidth]{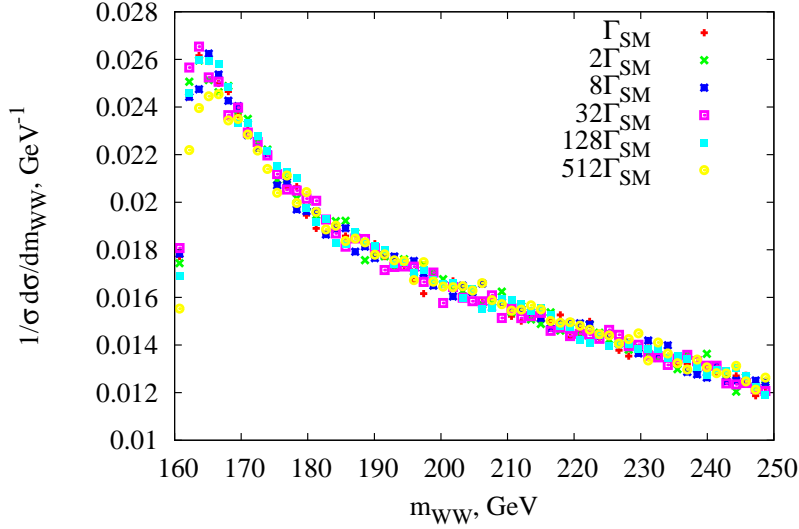} 
\end{center}
\caption{\label{width_WW} The dependence of the total cross section
  of $pp\to t\bar{t} W^{+}W^{-}$ at $\sqrt{s}=14$~TeV on the width of the
  Higgs boson for $m_{H}=150$~GeV, $\Gamma_{SM} = 0.01726$~GeV.
}
\end{figure}

Similarly one can  investigate another channel, $pp\to
t\bar{t}W^{+}W^{-}$. The dependence of the corresponding total cross
section on the Higgs boson mass and width is presented in
Figure~\ref{cs_WW}. The shapes of invariant mass $m_{WW}$ distribution
are plotted in Figures~\ref{WW} and~\ref{width_WW}. 

\begin{figure}[htb]
\begin{center}
\includegraphics[angle=0,width=0.80\columnwidth]{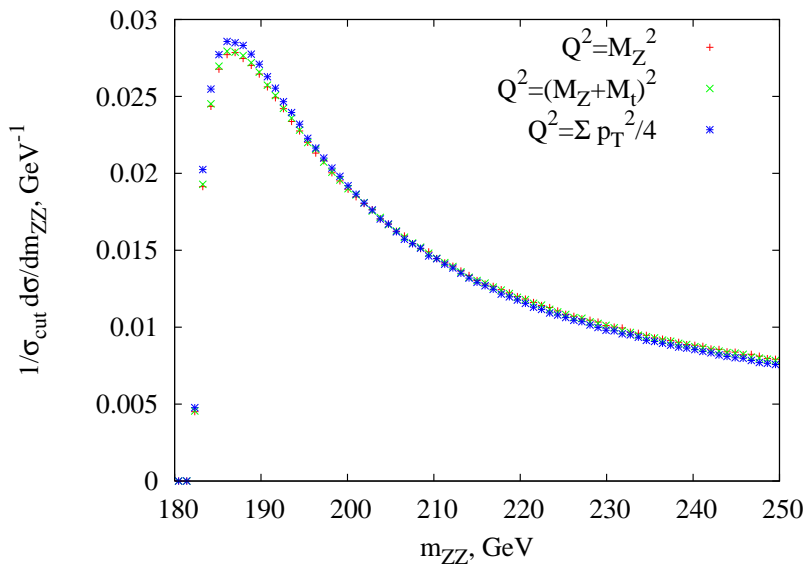} 
\end{center}
\caption{\label{nlo} The shapes of the invariant mass $m_{ZZ}$
  distribution in $pp\to t\bar{t} ZZ$ for different choices of
  renormalization scale at $\sqrt{s}=14$~TeV.
}
\end{figure}

Note in passing, that all our calculations above and below have been
done at the leading order in perturbative QCD. We leave a thorough
investigation of NLO corrections to future work, which is not
straightforward for any $2\to4$ process. In this paper we make
a crude estimate of these corrections and study their possible effect 
on the shape of $m_{ZZ}$ distribution and on the position of its
maximum. In all our numerical calculations we take  renormalization
scale for parton distribution functions to be $Q=M_Z$. Here we compare
the above results for total cross section and for $m_{ZZ}$
distribution to the results obtained with different choices of $Q$,
namely $Q=M_Z+M_t$ and $Q=\sqrt{\sum_{f}p^{2}_{T}}/2$, where sum is
taken over squared transverse momenta of all particles in the final
state. As an example, we consider the case of process $pp\to 
t\bar{t}ZZ$ with Standard model Higgs boson $m_{H}=170$~GeV,
$\sqrt{s}=14$~TeV. We find for the total cross section of this
processes 
\vskip 0.5cm
\centerline{\begin{tabular}{c|c}
$Q^2$ & $\sigma$,~fb \\ \hline
$M_Z^2$ & 1.44 \\
$\l M_Z+M_t\r^2$ & 0.85 \\ 
$\sum_{f}p^{2}_{T}/4$ & 1.03 
\end{tabular}
}
\vskip 0.5cm 
\noindent 
The behavior of the cross section with $Q$ is similar to that in
$t\bar t H$ final state, obtained in 
Refs.~\cite{Beenakker:2002nc,Dawson:2003zu}, 
with account of both leading order and next-to-leading order
corrections. Thus, one can expect quite a substantial NLO corrections
to the total signal cross section. However, as one can see from
Figure~\ref{nlo}, 
both the position of maximum and the shape of the $m_{ZZ}$
distribution, normalized to the cross section $\sigma_{cut}$,
  obtained by integrating the differential cross section
over the relevant interval $180~{\rm GeV}<m_{ZZ}<250~{\rm GeV}$,
remain intact with changing of the renormalization scale in the wide
range above. To make a numerical estimate of this uncertainty we
evaluate the expected number of events in 7 GeV width bins over
$m_{ZZ}$ for the three choices of $Q^2$, In a given bin the
corresponding three numbers differ by less than 5 percent. This
suggests that NLO corrections do not change significantly the shape of
this distribution and hence do not add to the uncertainties in the
Higgs boson parameters to be determined in the way we propose.  


\section{Invisible Higgs in $pp\to b\bar{b}ZZ$ and $pp\to
  b\bar{b}W^+W^-$ at LHC}
\label{Sec:bb}

It is worth noting that the total cross sections of the processes with
$t$-quarks considered in the previous section are of order of a few
fb, which requires high luminosity running of LHC to be of practical 
interest. The same is true for similar channels $pp\to b\bar{b}ZZ$ and
$pp\to b\bar{b}W^{+}W^{-}$ within the SM. Indeed, in this case the
Yukawa coupling of $b$-quarks to Higgs boson is quite small and the
diagrams with virtual Higgs boson are suppressed.  However, in many
promising extensions of the SM this Yukawa coupling increases, as it
takes place, for example, in the two-Higgs doublet model (2HD) and in  
Minimal Supersymmetric Standard Model (MSSM).  For illustrative
purposes we take the Yukawa coupling of $b$-quarks increased by a
factor $A=50$ with respect to the SM case (this refers to the value of
$\tan\beta=50$ in 2HD and MSSM).  Note, that the change of the
$b$-quark Yukawa coupling also yields a change of the Higgs boson
total width which we take into account accordingly. In these
modifications of the  SM with large value of $A$ the Higgs boson width 
$\Gamma_{mSM}$ is saturated by its decay into $b$-quarks, so that 
\begin{equation}
\label{modified-width}
\Gamma_{mSM}=A^2\cdot\Gamma_{SM}^{H\to b\bar b}\;,
\end{equation}
where $\Gamma_{SM}^{H\to b\bar b}$ is the SM Higgs boson decay rate
into b-quarks. 

\begin{figure}[htb]
\begin{center}
\begin{tabular}{ll}
\includegraphics[angle=0,width=0.50\columnwidth]{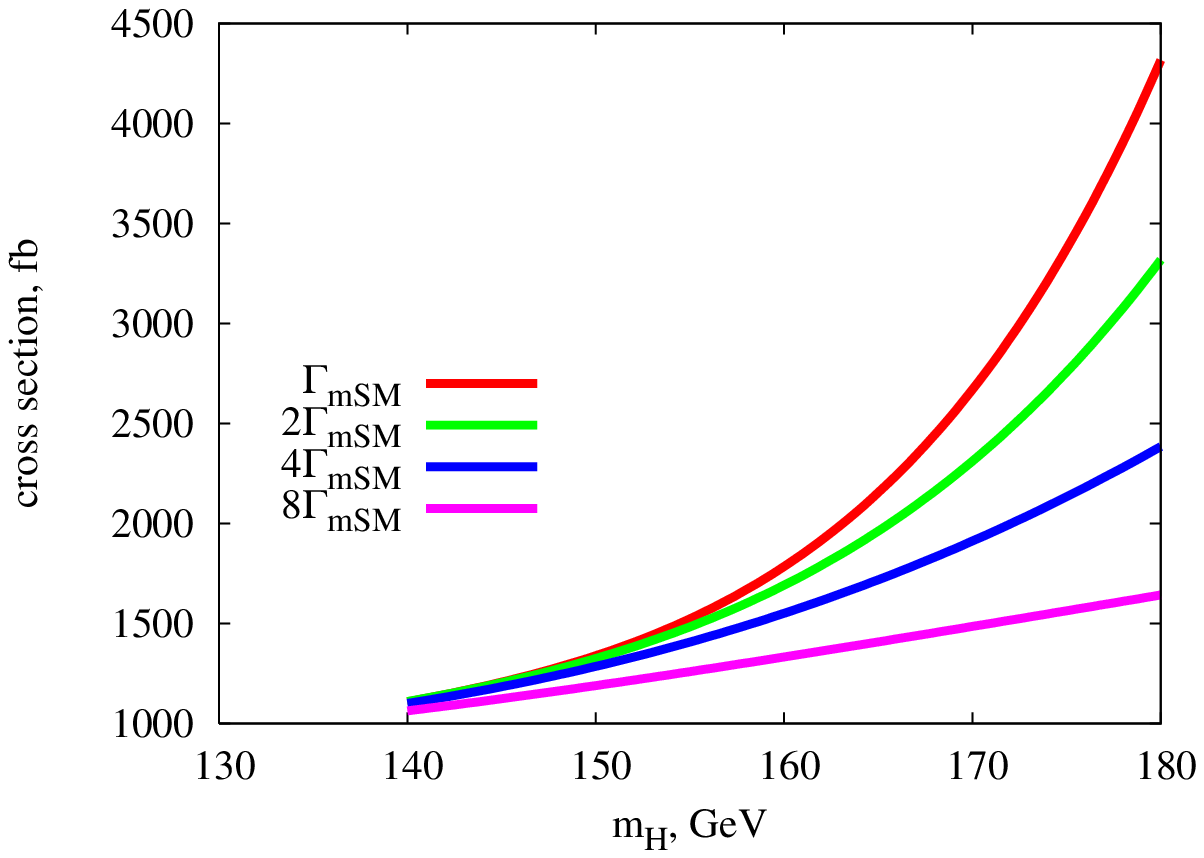} 
 &
\includegraphics[angle=0,width=0.50\columnwidth]{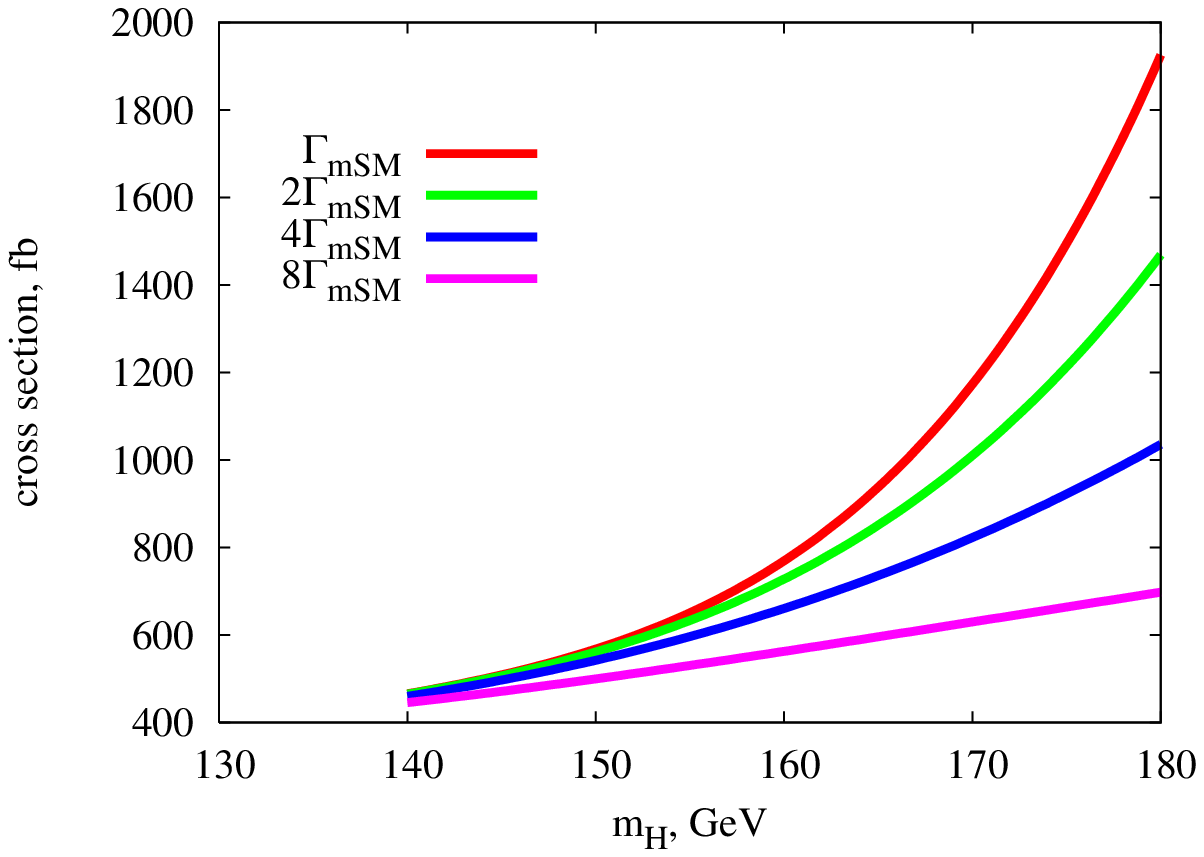} 
\end{tabular}
\end{center}
\caption{\label{cs_bbZZ} The dependence of the total cross section
  of $pp\to b\bar{b} ZZ$ at $\sqrt{s}=14$~TeV (left) and
  $\sqrt{s}=10$~TeV (right) on the mass of the Higgs boson in the
  modified Standard Model with b-Higgs coupling enhanced by factor
  $A=50$; $\Gamma_{mSM}$ is given by Eq.~\eqref{modified-width}.  
}
\end{figure}

\begin{figure}[htb]
\begin{center}
\begin{tabular}{ll}
\includegraphics[angle=0,width=0.50\columnwidth]{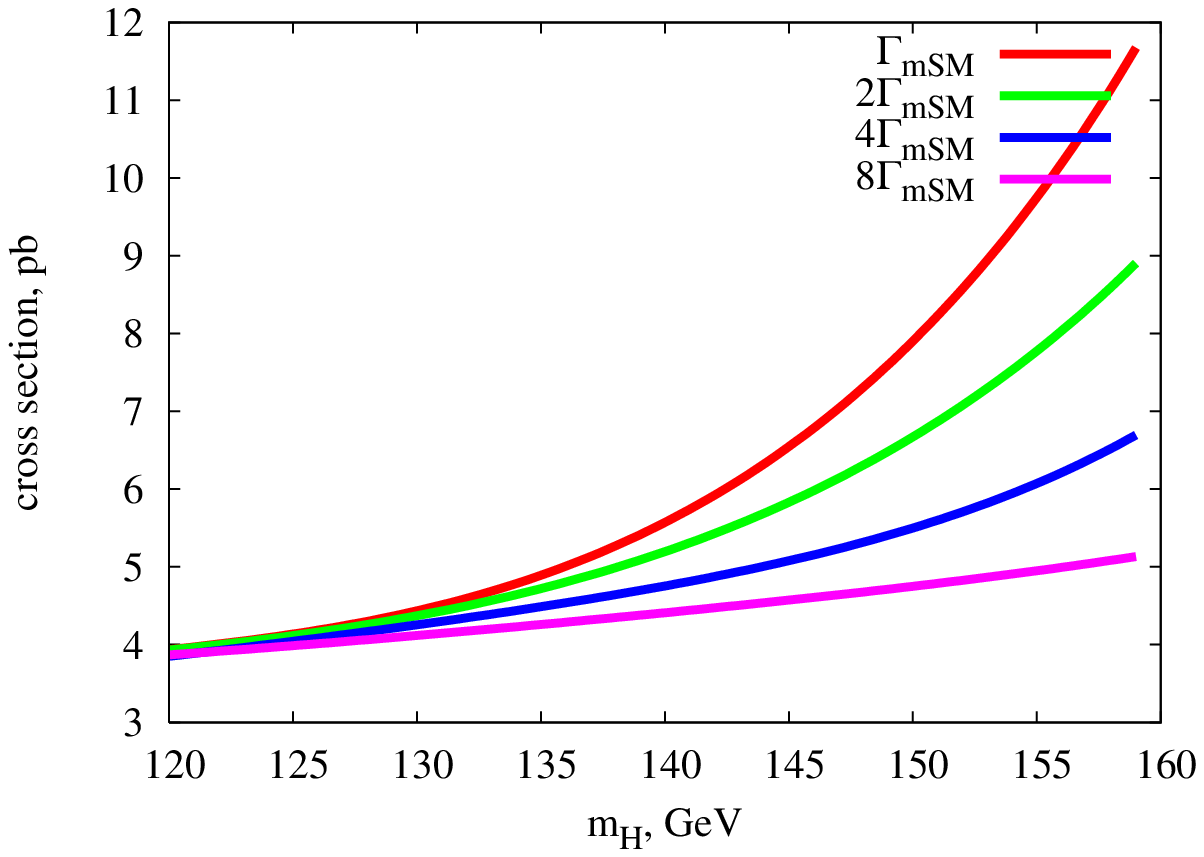} 
 &
\includegraphics[angle=0,width=0.50\columnwidth]{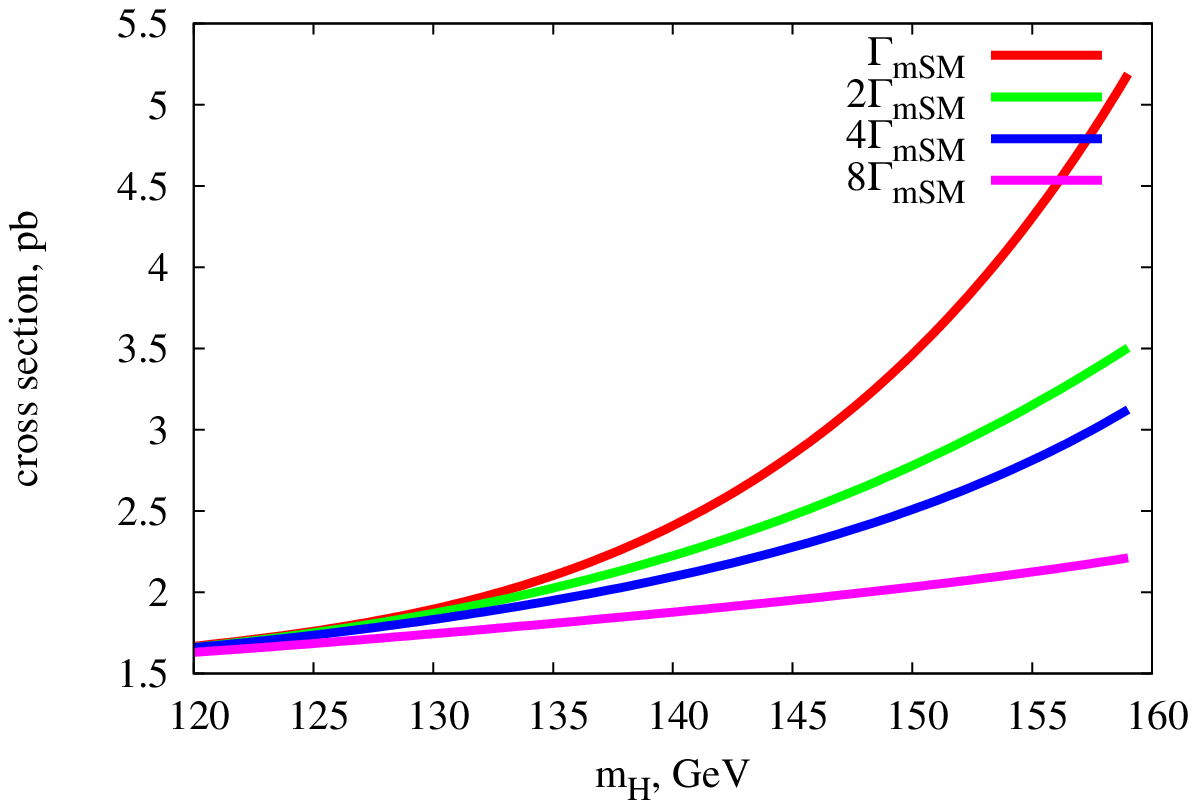} 
\end{tabular}
\end{center}
\caption{\label{cs_bbWW} The dependence of the total cross section
  of $pp\to b\bar{b} W^{+}W^{-}$ at $\sqrt{s}=14$~TeV (left panel) and 
  $\sqrt{s}=10$~TeV (right panel) on the mass of the Higgs boson in the 
  modified Standard Model with b-Higgs coupling enhanced by factor
  $A=50$; $\Gamma_{mSM}$ is given by Eq.~\eqref{modified-width}. We
  exclude the following part of the phase space: 
  $159.3~{\rm GeV}<M_{bW^{+}}<189.3$~GeV, $159.3~{\rm
  GeV}<M_{\bar{b}W^{-}}<189.3$~GeV. 
}
\end{figure}

For the processes with $b$-quarks we exclude from considerations the
following regions of the phase space of the final state:
$159.3~{\rm GeV}<m_{bW^{-}}<189.3$~GeV and $159.3~{\rm
  GeV}<m_{\bar{b}W^{+}}<189.3$~GeV because in these regions the cross
section is saturated by top-quark production and the interesting
effects get obscured. The 
corresponding total cross sections are given in Figure~\ref{cs_bbZZ}
for $b\bar{b}ZZ$ final state and in Figure~\ref{cs_bbWW} for
$b\bar{b}W^{+}W^{-}$.

\begin{figure}[htb]
\includegraphics[angle=0,width=0.90\columnwidth]{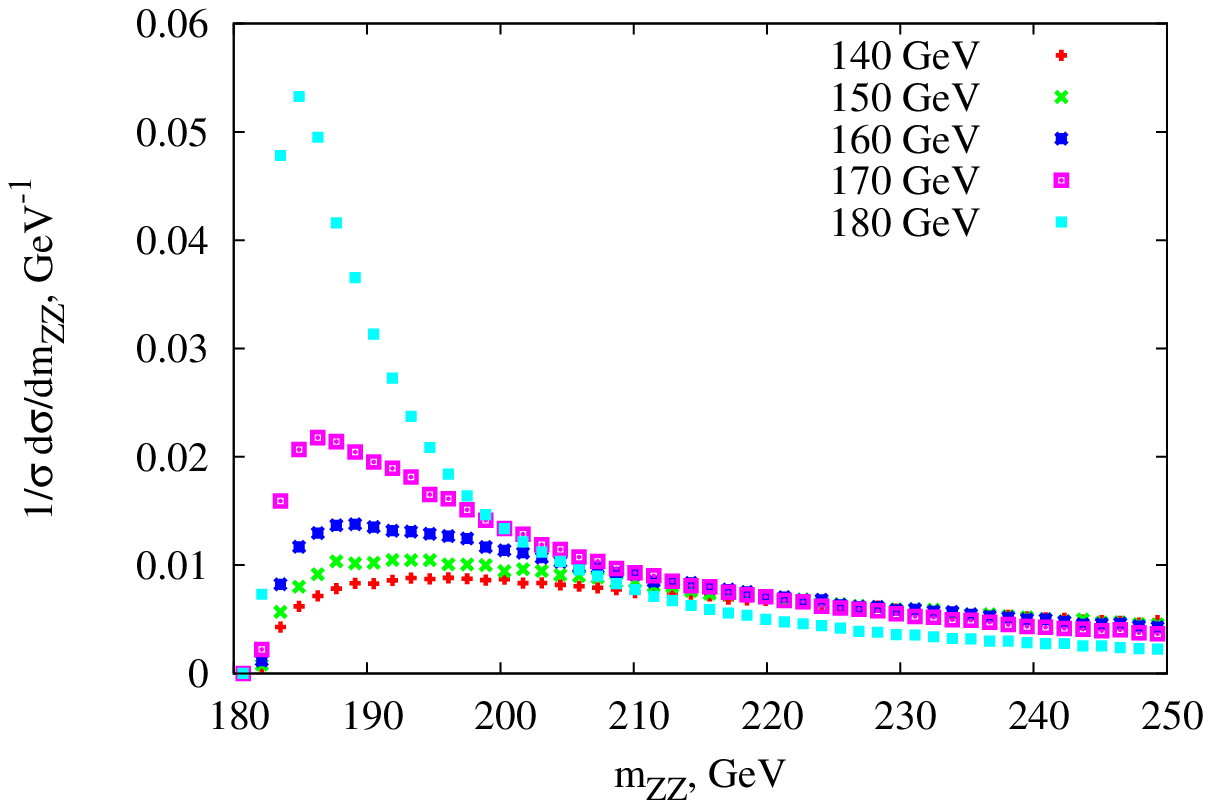} 
\includegraphics[angle=0,width=0.90\columnwidth]{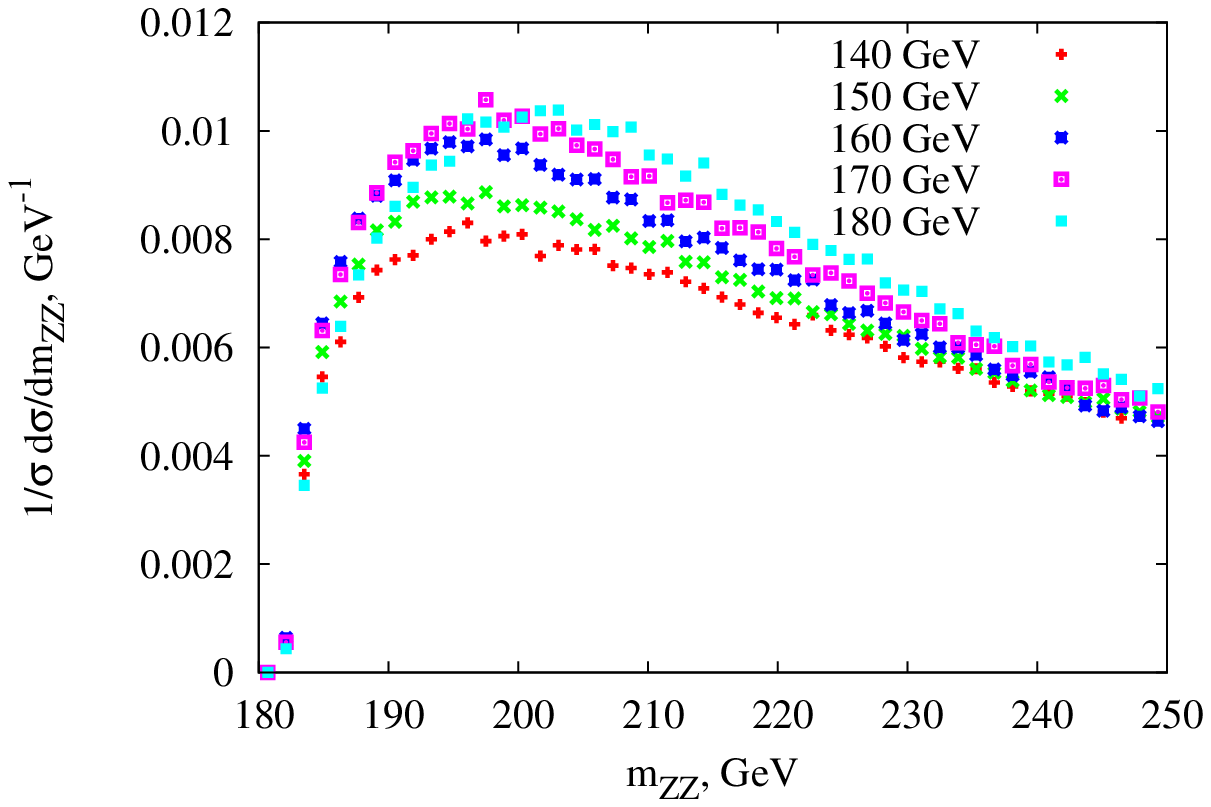} 
\caption{\label{bb_ZZ} The invariant mass $m_{ZZ}$ distribution in
  the $pp\to b\bar{b} ZZ$ process for several values of the Higgs
  boson mass for modified Standard Model Higgs boson width
  $\Gamma_{mSM}$ given by Eq.~\eqref{modified-width} (upper panel) and
  for the width 8 times larger due to invisible decay mode (lower
  panel).
}
\end{figure}

\begin{figure}[htb]
\includegraphics[angle=0,width=0.9\columnwidth]{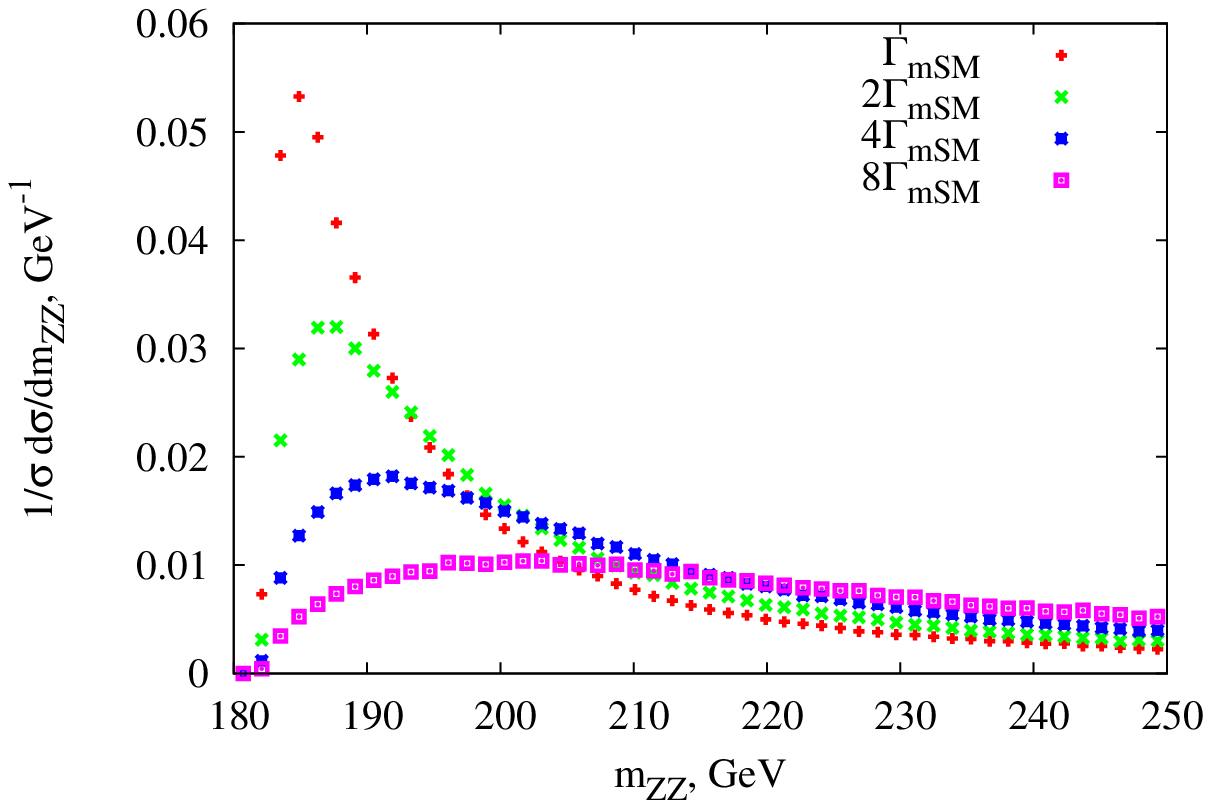} 
\includegraphics[angle=0,width=0.9\columnwidth]{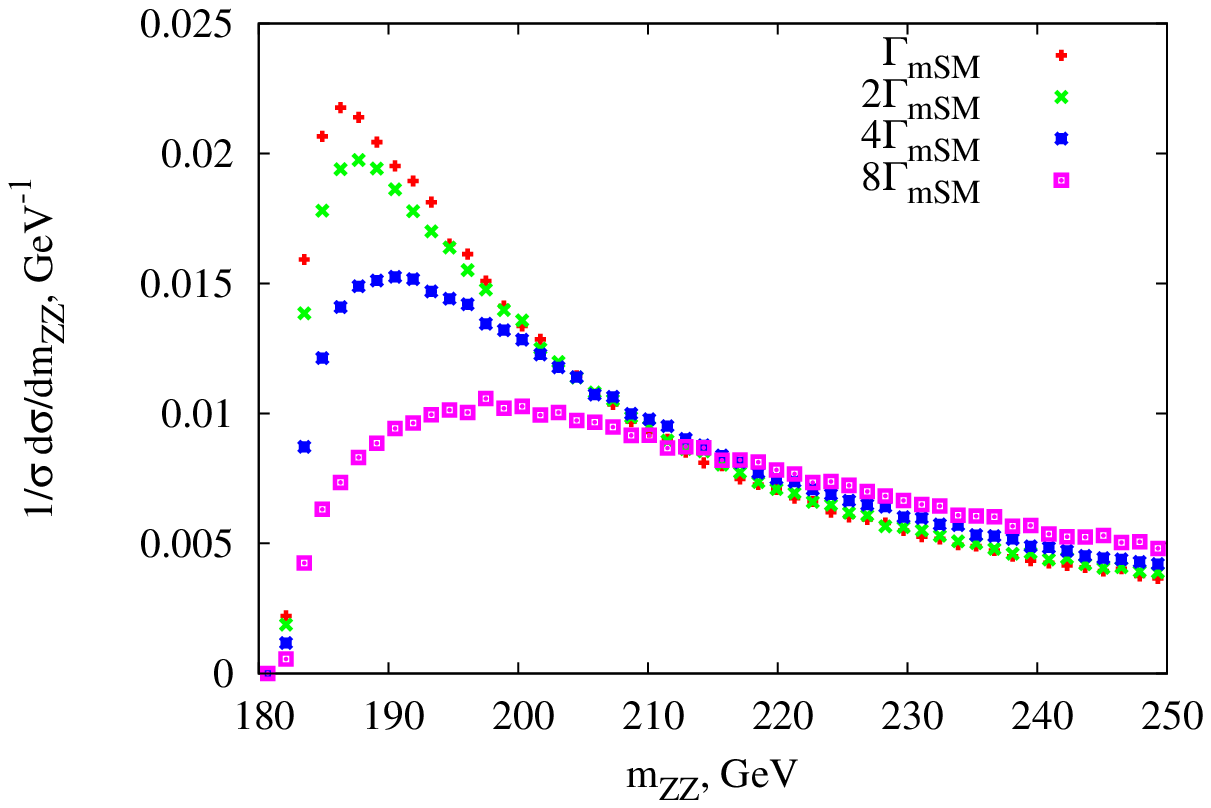} 
\caption{\label{width_bb_ZZ} The dependence of the total cross section
  of $pp\to b\bar{b} ZZ$ on the width of the Higgs boson for
  $m_{H}=180$~GeV, $\Gamma_{mSM} = 9.04$~GeV (upper panel) and
  $m_{H}=170$~GeV, $\Gamma_{mSM} = 8.41$~GeV (lower 
  panel). $\Gamma_{mSM}$ is given by Eq.~\eqref{modified-width} 
with factor $A=50$. 
}
\end{figure}

For the processes with $b$-quarks we observe qualitatively similar  
dependence of the $m_{ZZ}$ distribution on the Higgs boson mass; 
remarkably, this distribution depends also on the Higgs boson width,
see Figures~\ref{bb_ZZ} and~\ref{width_bb_ZZ}. However, both the shape
of $m_{ZZ}$ distribution and the position of its maximum are much more
sensitive to $m_H$ and $\Gamma_H$ as compared to the case of
$t$-quarks. So, to obtain the Higgs boson width and mass one should
make two-parametric analysis of $m_{ZZ}$ distribution and total cross
section. At the same time, with the same collected statistics one can
expect to achieve higher accuracy in measurements of the Higgs boson
mass and width, than in the channels with $t$-quarks. Of course
  this is true only for large enough values for the constant A.

\section{Study of partonic scattering $b\bar{b}\to ZZ$}
\label{Sec:analytic}

This example demonstrates the main effect under discussion by simple
analytical calculation of the three diagrams --- $t$- and $u$-channel
$b$-quark exchange  and $s$-channel Higgs boson exchange ---
contributing to $bb\to ZZ$. 

The point is that processes $b\bar{b}\to ZZ$ and $gg\to b\bar{b}ZZ$
are closely related. It goes as follows.  The b-quark mass is rather
small as compared to the discussed range of the invariant mass
$M_{ZZ}$. It leads to the well known enhancement of the contribution
to the process $gg\to b\bar{b}ZZ$ from the kinematic region with two
very forward b-quarks by the factor $log(M_{ZZ}^2/m_b^2)$. This fact
allows one to resum the log factors by introducing the b-quark
distribution functions in protons and study the process
$b\bar{b}\to ZZ$ with the b-quarks in the initial states (see,
\cite{Maltoni:2003pn, Boos:2003yi, Campbell:2004pu} and references
therein).

Total partonic cross section of the process $b\bar{b} \to ZZ$ is obtained
with the help of symbolic part of CompHEP and its complete expression
is given in Appendix~A. Here we present only the zero and the first
order terms in $1/s$ (see notations in the Appendix~\ref{appendixA}) 
\bea
\sigma_{s} = \frac{\pi\alpha}{6s_{W}^{4}c_{W}^{4}M_{Z}^{2}s^2}\left\{
\frac{1}{2}(l-r)^4M_{b}^2s^2 +
M_{Z}^4s\left[2(l^4+r^4)\left(\log{\frac{s}{M_{b}^2}}-1\right)
\right.\right.\nonumber\\
\left.\left.
-4(l^4-l^3r-lr^3+r^4)\right]
+M_{b}^2M_{Z}^2s\left[2lr^2(l-r)-(l-r)^4\right]
\right.\nonumber\\
\left.
-(l-r)^4M_{b}^4s\left(\log{\frac{s}{M_{b}^2}}+1\right)
+\frac{1}{4}A(l-r)^2\left]
-M_{b}^2s^2+4M_{b}^2M_{Z}^2s
\right.\right.\\
\left.\left.
+M_{b}^4s\left(\frac{1}{2}\log{\frac{s}{M_{b}^2}}+2\right)
-(m_{H}^2-\Gamma_{H}^2)M_{b}^2s\right]
+\frac{1}{16}A^2\left[
\frac{1}{2}M_{b}^2s^2
\right.\right.\nonumber\\
\left.\left.
-3M_{b}^2M_{Z}^2s-3M_{b}^4s+\frac{1}{2}(2m_{H}^2-\Gamma_{H}^2)M_{b}^2s
\right]
\right\}\nonumber
\eea
The partonic cross section should be converted with parton
distribution functions of $b$-quarks in proton $f_{b}(x, Q^2)$ and 
$f_{\bar{b}}(x, Q^2)$ (where $x$ is the fraction of proton momentum
carried by a parton and $Q$ is the characteristic QCD factorization
scale taken here equal to be the Higgs mass). This is done with the
CompHEP using the b-quark parton distribution functions as given by
CTEQ6L1~\cite{Pumplin:2002vw}. 

\begin{figure}[htb]
\includegraphics[angle=0,width=0.90\columnwidth]{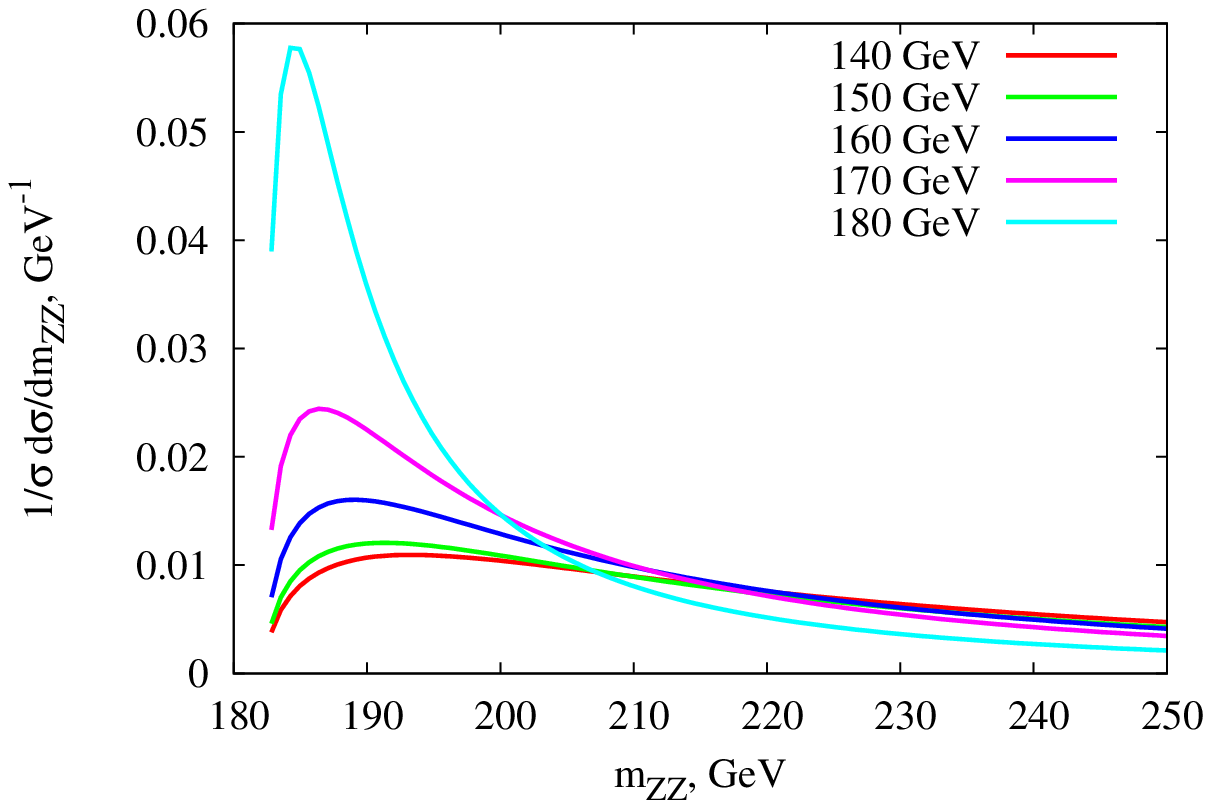} 
\includegraphics[angle=0,width=0.90\columnwidth]{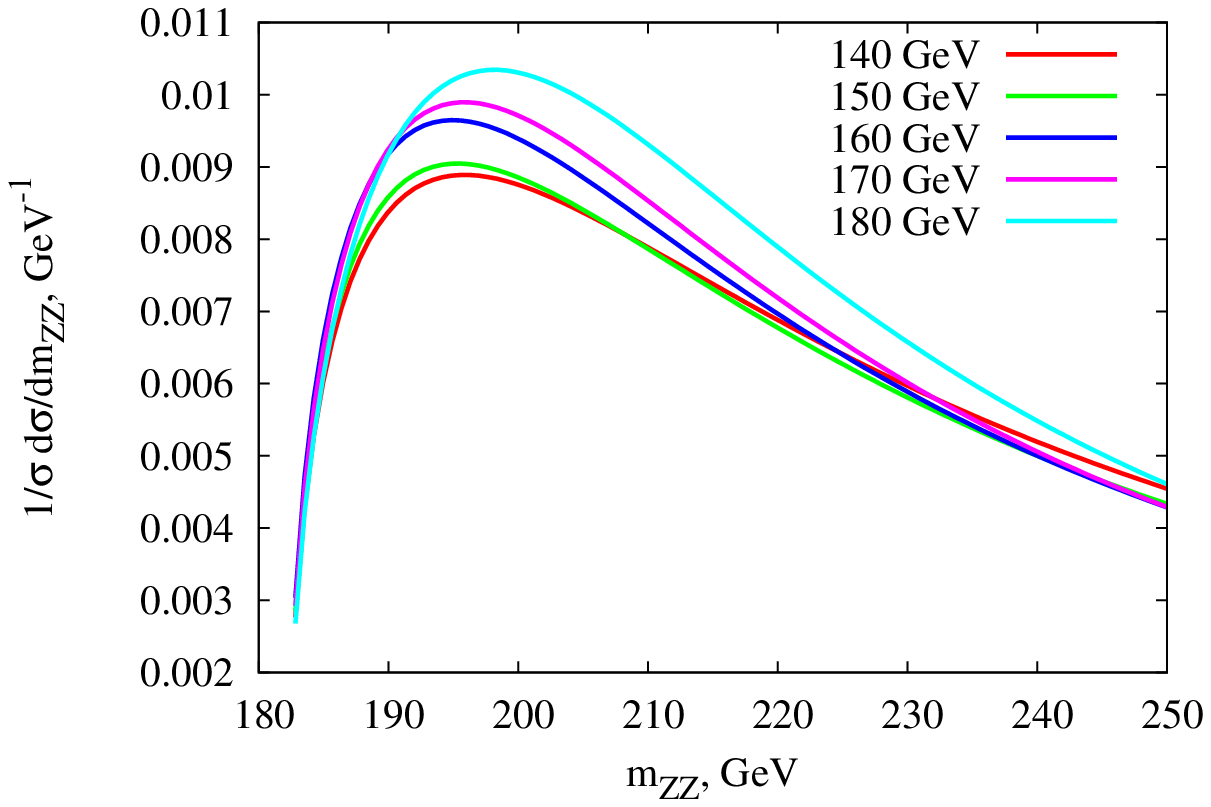} 
\caption{\label{formula_bb_ZZ} The invariant mass $m_{ZZ}$
  distribution for the $b\bar{b}\to ZZ$ process for several values of
  the Higgs boson mass in the modified SM with enhanced b-Higgs
  coupling by factor $A=50$, see Eq.~\eqref{modified-width}, for the
  Higgs boson width $\Gamma_{mSM}$ (upper panel) and for the width 8
  times larger (lower panel) calculated with analytical 
  formula~(\ref{formula}). 
}
\end{figure}

\begin{figure}[htb]
\includegraphics[angle=0,width=0.9\columnwidth]{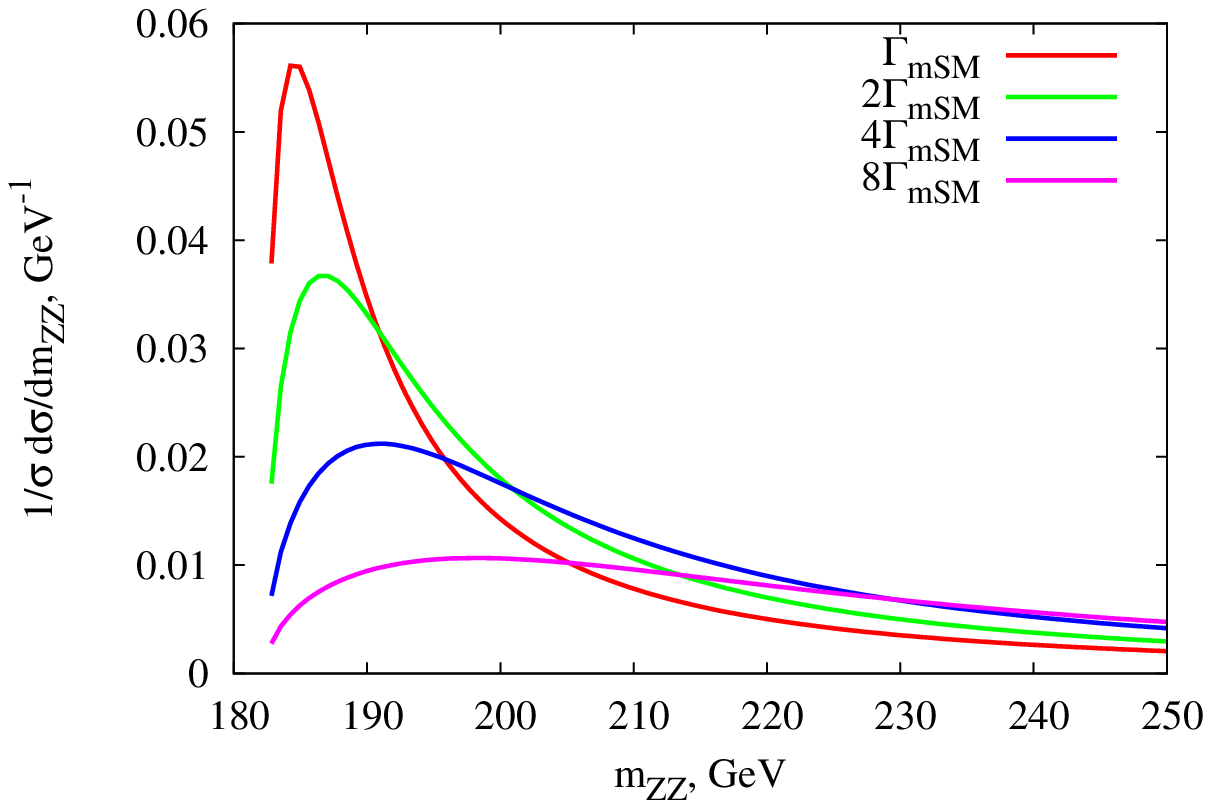} 
\includegraphics[angle=0,width=0.9\columnwidth]{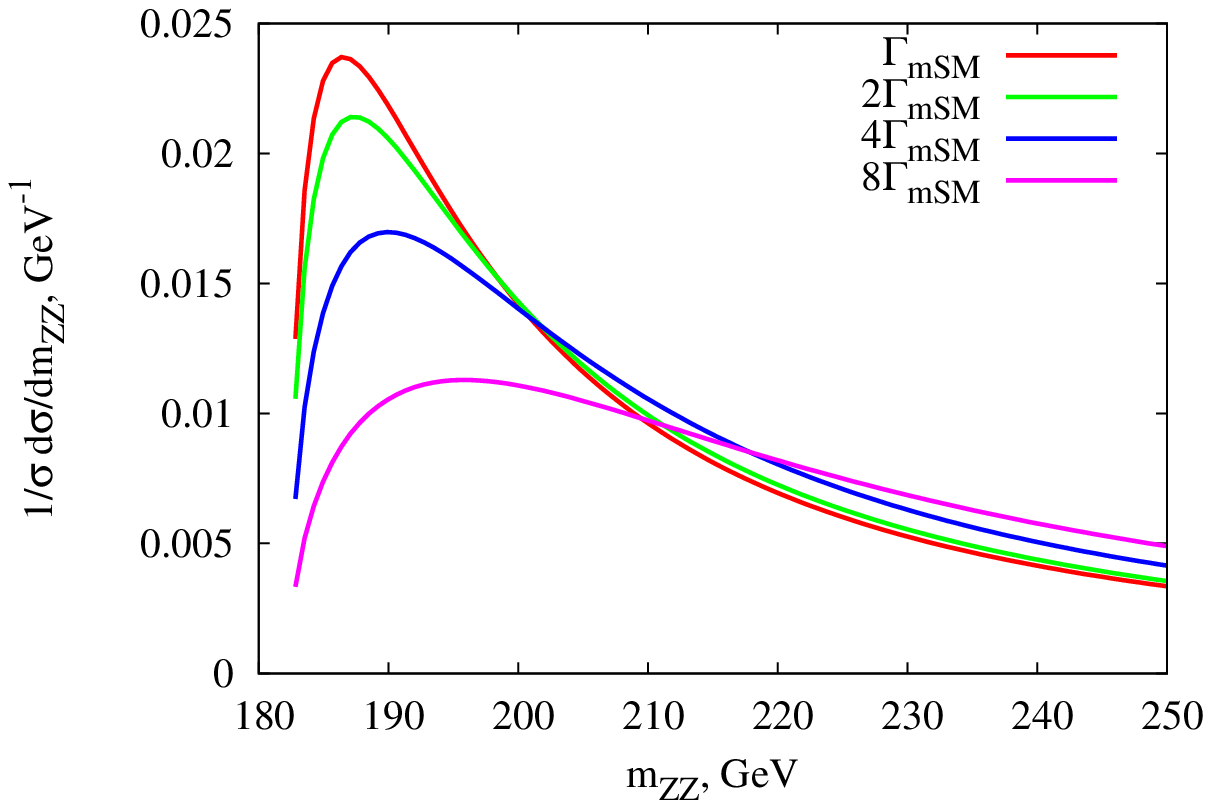} 
\caption{\label{formula_width_bb_ZZ} The dependence of the total cross
  section $b\bar{b} \to ZZ$ on the width of the Higgs boson in the
  modified SM with enhanced b-Higgs coupling by factor $A=50$, see
  Eq.~\eqref{modified-width}, for $m_{H}=180$~GeV, $\Gamma_{H} =
  9.04$~GeV (upper panel) and $m_{H}=170$~GeV, $\Gamma_{H} = 8.41$~GeV
  (lower panel) calculated with analytical formula~(\ref{formula}). 
}
\end{figure}

Then the dependence of the invariant mass $m_{ZZ}$ on mass and width
of the Higgs boson for $pp\to b\bar{b}\to ZZ$ is presented in 
Figures~\ref{formula_bb_ZZ} and~\ref{formula_width_bb_ZZ}. By
comparing them with Figs.~\ref{bb_ZZ} and~\ref{width_bb_ZZ} we
see a nice agreement. 


One can understand the behavior of the distribution over $m_{ZZ}$ in
the process $pp\to b\bar{b}ZZ$ from rather general arguments. Let us 
represent the corresponding matrix element as a sum $M=M_{1}+M_{2}$
where the second term refers to the diagrams describing $Z$-boson pair
production via virtual Higgs boson (so, it has the form $M_{2}\sim
\frac{1}{m_{ZZ}^{2} - m_{H}^{2}+im_{ZZ}\Gamma_{H}}$), while $M_{1}$
includes all other relevant contributions. Then, integrating partly
over the phase space one arrives the integral over invariant
mass $m_{ZZ}$, 
\be
\label{simple-illustration}
\begin{split}
\sigma_s = \frac{1}{8\pi}\int dm^{2}_{ZZ}
\sqrt{1 - \frac{4m_{Z}^2}{m_{ZZ}^2}} \left\{
f_{11}(m_{ZZ}) +
\frac{2f_{12}(m_{ZZ})(m_{ZZ}^2-m_{H}^2)} {(m_{ZZ}^2-m_{H}^2)^2 +
  \Gamma_{H}^{2}m_{ZZ}^2} 
\right.\\
\left.
+ \frac{f_{22}(m_{ZZ})}{(m_{ZZ}^2-m_{H}^2)^2
  + \Gamma_{H}^{2}m_{ZZ}^2}
\right\}\;.   
\end{split}
\ee
Here the square root term comes from two-particle phase space and we
omit the dependence of functions $f_{11}, f_{12}, f_{22}$ on
$s$. These functions of $m_{ZZ}$ vary rather mildly in a quite broad
interval about $m_H$. Treating them as constants one checks
that the integrand in Eq.~\eqref{simple-illustration} fits the
$m_{ZZ}$ invariant mass distribution presented in
Section~\ref{Sec:bb}. Similar  reasoning is valid for other channels
under discussion. Thus the behavior of $m_{ZZ}$ distribution observed
in this work reflects the Breit-Wigner structure of the virtual Higgs
boson contribution. 

In the vicinity of the Higgs mass $m_{H}$ the main contribution to the
shape of $m_{ZZ}$ invariant mass distribution as a functions of
$m_{ZZ}$ comes from the third term in Eq.~\eqref{simple-illustration}
containing $f_{22}(m_{ZZ})$. Hence if properly normalized this
distribution approximately does not depend on function
$f_{22}(m_{ZZ})$ at all (recall we assume that $f_{22}(m_{ZZ})$ is a
constant in the some interval near $m_{H}$). In this way one can argue
that $m_{ZZ}$ invariant mass distribution is not significantly
affected by NLO corrections, confirming conclusions of numerical
studies in Section 2.

\section{Conclusions}
\label{Sec:conclusion}
We have performed a phenomenological analysis of the Higgs boson
manifestation at the LHC in the processes of heavy quarks (top and
bottom) pair production in association with the electroweak  bosons
$W^+W^-$ and $ZZ$ in the framework of the models where the Higgs boson
decay branching rations for SM is practically small (e.g., like in
Higgs portal models). In this case one expects that the Higgs boson
observation as a resonance is problematic since all the SM decay modes
are very much suppressed. However, the Higgs boson should play its
role in unitarization of the weak amplitudes behavior, and this
property could be exploited to search for the Higgs boson
manifestation. We performed complete tree level computations of the
relevant $2\to4$ sub-processes numerically and illustrated the main
observations by simple symbolic formulas. Numerical analysis is
presented for the case of LHC energy 
of 14 TeV (particular results are given for the energy of 10 TeV for
comparison). The 
total cross sections of the processes $pp\to t\bar{t}VV$, where $V$ is
$Z$- or $W$-boson, are sensitive to a variation of both the Higgs mass
and width in the Higgs mass region somewhat less than the mass of
vector boson pair ($130-140~{\rm GeV} < M_H < M_{VV}$). The invariant mass
distribution of vector boson pairs also exhibits strong dependence on
the Higgs boson mass and width. The processes with the b-quarks $pp\to
b\bar{b}VV$ 
obviously have too small rate for the case of the SM Higgs coupling,
however the processes might be interesting in models where the b-quark
coupling to the Higgs boson is significantly increased, for
example, in MSSM like models with large $\tan{\beta}$. In case of
$pp\to b\bar{b}W^+W^-$ one should cut out the phase space region of W
and b invariant mass around the top quark mass to remove a very large
top pair contribution and to observe the effect of cross section and
distribution dependence on the Higgs mass and width. We have estimated
an impact of possible QCD corrections on observed dependence by
performing computations at various QCD renormalization
scales. In accordance with known NLO computations for the Higgs production
in association with heavy quarks the cross sections and distributions
get 
corrections, however, the observed dependence on the Higgs mass and width
is not practically affected. Complete tree level computations of the
processes with 4 final state particles allows to take into account all
the irreducible backgrounds. However, the influence of reducible
backgrounds and effects of detector finite resolutions are needed to
be taken into account for more realistic analysis. In particular, if
one takes into account subsequent top quark, W- and Z-bosons decays
one may study off-shell production and hope to pin down the
sensitivity region for the Higgs mass below $130-140$~GeV. However
this looks problematic since rather small rate for considered on-shell
production processes will be even smaller in this case.

{\bf Acknowledgments. }  The work was supported by Russian Ministry of
Education and Science under state contract 02.740.11.0244. The work of
E.B. was  also supported by the grant of Russian Ministry of Education 
and Science NS-4142.2010.2, RFBR grants 08-02-91002-CERN$\_$a and
08-02-92499-CNRSL$\_$a. D.G. thanks the organizers of the long-term
workshop in Yukawa Institute YITP-T-10-01 for hospitality.  The work
of D.G. and S.D. was supported in part by the Russian Foundation for
Basic Research (grants 08-02-00473a), by the grants of the President
of the Russian Federation NS-5525.2010.2 and by FAE program
(government contract $\Pi$520). The work of S.D. was also supported by
the grants of the President of the Russian Federation MK-4317.2009.2,
by FAE program (government contract $\Pi$2598). Numerical part of the
work was performed on the Computational cluster of the Theoretical
Division of INR RAS. 

\appendix 
\section{Formula for cross section $b\bar{b}\to ZZ$}
\label{appendixA} 

Let us introduce following notations
\be
s_W=\sin{\theta_{W}},\;\;\;
l = T_{3} - Qs_{W}^{2} \equiv -\frac{1}{2} +
\frac{1}{3}s_{W}^{2},\;\;\;
r = -Qs_{W}^{2} \equiv \frac{1}{3}s_{W}^{2},
\ee

\be
\begin{split}
A_s(s) = \sqrt{s-4M_{b}^{2}}\sqrt{s-4M_{Z}^2},\;\;\;
U(s) & = s+A_s(s)-2M_{Z}^{2}\;,\\
D(s) & = s-A_s(s)-2M_{Z}^{2}\;. 
\end{split}
\ee
\bea
\sigma_{s} = \frac{\pi\alpha}{6s_{W}^{4}c_{W}^{4}M_{Z}^{2}s^2}\left\{ 
\l\frac{1}{D}-\frac{1}{U}\r\left[-8 \l l^4+r^4\r M_{Z}^{8}
+8M_{Z}^{6}M_{b}^{2}\l l^4-5l^3r+2l^2r^2-5lr^3 
\right.\right.\right.\nonumber\\
\left.\left.\left.
+r^4\r 
-2M_{Z}^{4}M_{b}^{4}\l l^4-20l^3r+54l^2r^2-20lr^3+r^4\r 
-
8M_{b}^{2}M_{Z}^{4}s\l l^4-l^3r-lr^3+r^4\r\right] 
\right. \nonumber \\
\left.
+\frac{1}{2M_{Z}^{2}-s}\log{\frac{U}{D}}\left[-8\l l^4+r^4\r M_{Z}^{8} +
  4M_{Z}^{6}M_{b}^{2}\l 6l^4-10l^3r+17l^2r^2-11lr^3+6r^4\r 
\right.\right.\nonumber\\
\left.\left.
-2M_{Z}^4M_{b}^{4}\l 3l^4-12l^3r+2l^2r^2-12lr^3+3r^4\r
-8M_{b}^{8}\l l^4+r^4\r 
+4M_{Z}^{4}M_{b}^{2}s\l l^4+2l^3r
\right.\right.\right.\nonumber\\
\left.\left.\left.
-11l^2r^2+3lr^3+r^4\r 
+ \frac{3}{2}M_{b}^{4}M_{Z}^{2}s\l l^4-8l^3r+14l^2r^2-8lr^3+r^4\r 
\right.\right.\nonumber\\
\left.\left.
+4M_{b}^{6}s\l l^4+r^4\r + M_{b}^{4}s^2\l l-r\r^4 -
2\l l^4+r^4\r M_{Z}^{4}s^{2}\right]
\right. \label{formula}\\
\left.
+A_s(s)\left[-2\l l^4+r^4\r M_{Z}^{4} +
  M_{Z}^{2}M_{b}^{2}\l 2l^2r^2-2lr^3\r +
  \frac{1}{2}(l-r)^4M_{b}^{2}s\right] 
\right.\nonumber\\
\left.
+\frac{AM_{b}^{2}\l s-m_{H}^{2}\r}{\l s-m_{H}^{2}\r^2 + s\Gamma_{H}^{2}}
\left[A_s(s)\l-\frac{1}{4}(l-r)^2s^2 
  + \frac{1}{2}\l l-r\r^2M_{Z}^{2}s - 2\l l^2+r^2\r M_{Z}^{4}\r
\right.\right.\nonumber\\
\left.\left.
+ \log{\frac{U}{D}}\left(-4(l-r)^2M_{Z}^{4}M_{b}^{2} 
+\frac{1}{2}\l l-r\r^2M_{b}^{2}s^2
-2\l l-r\r ^2M_{b}^{2}M_{Z}^{2}s - 2\l l^2+r^2\r M_{Z}^{6}
\right.\right.\right.\nonumber\\
\left.\left.\left.
+ 4\l l^2-lr+r^2\r M_{Z}^{4}s\right)\right]
+\frac{A^2M_{b}^{2}A_s(s)}{32\l\l s-m_{H}^{2}\r^2+s\Gamma_{H}^{2}\r
}\l s-4M_{b}^{2}\r\l 12M_{Z}^4-4M_{Z}^2s + s^2\r\right\}\;. \nonumber
\eea

Here $s=(p_1+p_2)^2$ where $p_1$ and $p_2$ are the momenta of $b$ and
$\bar{b}$ quarks, respectively. One can check that the leading terms
in $s$ (which 
lead to constant cross section at high energies, thus violating
unitarity) cancel at $A=1$, otherwise they should be canceled by terms
coming from other non-SM particles. 

We recall that the width of the Higgs boson $\Gamma_{H}$ depends on
the amplification factor $A$ as well. 

\end{document}